\documentclass[aps,prc,twocolumn,groupedaddress]{revtex4}

\usepackage{graphicx}
\usepackage{longtable}
\usepackage{subfigure}
\def\vc#1{\mbox{\boldmath $#1$}}
\begin{document}

%\preprint{}

\title{Resonance states in $^{12}$C and $\alpha$-particle condensation}


\author{Y.~Funaki$^1$, A.~Tohsaki$^2$, H.~Horiuchi$^1$, P.~Schuck$^3$, 
and G.~R\"opke$^4$}
%\thanks{}
\affiliation{
$^1$ Department of Physics, Kyoto University, Kyoto 606-8502, Japan \\
$^2$ Department of Fine Materials Engineering, Shinshu University, Ueda 386-8567, Japan \\
$^3$ Institut de Physique Nucl\'eaire, 91406 Orsay Cedex, France \\
$^4$ FB Physik, Universit\"at Rostock, D-18051 Rostock, Germany
}
\date{\today}

\begin{abstract}
The states with $J^\pi=0^+$, $2^+$, and $4^+$ of $^{12}$C with excitation energies less than about 15 MeV are investigated with the alpha condensate wave function with spatial deformation and by using the method of ACCC (analytic continuation in the coupling constant) which is necessary for a proper treatment of resonance states. The calculated energy and width of the recently observed $2_2^+$ state are found to be well reproduced. The obtained $2_2^+$ wave function has a large overlap with a single condensate wave function of $3\alpha$ gas-like structure. The density distribution is shown to be almost the same as that of the $0_2^+$ state that is regarded as a $3\alpha$ Bose-condensed state, if the energy of the $2_2^+$ state is scaled down to the same value as the one of the $0_2^+$ state. Furthermore, the kinetic energy, nuclear interaction energy, and Coulomb interaction energy of the calculated $2_2^+$ state are shown to be very similar to those of the $0_2^+$ state. We conclude that the $2_2^+$ state has a structure similar to the $0_2^+$ state of Bose-condensate character with a dilute $3\alpha$ gas-like structure. In addition the resonance states, $0_3^+$, $0_4^+$, $4_2^+$, are also discussed.

\end{abstract}

%\keywords{}


\maketitle

\section{Introduction}\label{intro}
In our previous paper \cite{thsr}, we made the strong suggestion that $^{12}$C, $^{16}$O, and other self-conjugate $4n$ nuclei should have a gas-like structure of alpha particles in their excited states near the $n\alpha$ breakup threshold, and that this structure could be regarded as a kind of Bose-Einstein condensate, though only a few Bosons are concerned in nuclear systems. The $0_2^+$ state of $^{12}$C, called the Hoyle state and observed at 0.38 MeV above the $3\alpha$ threshold, and the $0_5^+$ state of $^{16}$O, observed at 0.44 MeV below the $4\alpha$ threshold, were indentified as candidates of the three-alpha condensed state and a possible candidate of the four-alpha condensed state, respectively by an investigation where the following type of wave function was adopted:
\begin{equation}
 \Phi_{n\alpha} \propto {\cal A}\ \Big[ \prod_{i=1}^n \exp \Bigl(-2 \frac{X_i^2}{B^2} \Bigr) \phi(\alpha_i) \Big]. \label{eq:1}
\end{equation}
Here $\phi(\alpha_i) \propto \exp [ - (1/2b^2) \sum_j^4 ({\vc r}_{j(i)} - {\vc X}_i )^2 ]$ and ${\vc X}_i = \sum_j^4 {\vc r}_{j(i)} / 4$, with $j(i)\equiv 4(i-1)+j$, are the internal wave functions and center-of-mass coordinates of the $i$-th alpha cluster, respectively. This wave function has the characteristic feature that all center-of-mass motions of the constituent alpha clusters occupy the same spherical $S$-wave orbit, $\exp (-2X^2 / B^2 )$. It also distinguishes itself from other $\alpha$-cluster models by the fact that the alpha clusters do not have any definite geometric configuration, but are allowed to move freely with small inter-alpha correlations or interactions, forming a dilute gas-like structure. It should be noted that the above mentioned characteristics can be put forward only as far as the variational parameter $B$ takes a sufficiently large value so that the action of the antisymmetrizer, ${\cal A}$, is weak and almost negligible. 

It is to be noted that the excited state of a self-conjugate $4n$ nucleus which has a well-developed $n\alpha$ cluster structure around the $n\alpha$ breakup threshold has almost zero inter-alpha binding energy and is only held together by the Coulomb barrier. Since $\alpha$ particles are Bosons, the gas-like state with almost zero inter-alpha binding energy is naturally considered to have Bose-condensed structure (see also \cite{yamada}).

For the $n$=3 case, i.e. $^{12}$C, detailed analysis was made by several authors about a quarter century ago, and they revealed the fact that several excited states observed in the vicinity of the $3\alpha$ particle threshold have an apparent $3\alpha$ cluster structure \cite{carbon,uegaki,kamimura,hori}. Especially, complete $3\alpha$ calculations with fully microscopic models were performed by two groups, Uegaki {\it et al.} \cite{uegaki}, and Kamimura {\it et al.} \cite{kamimura}. The calculated binding energy of the $0_2^+$ state was not much different from the experimental value which is 0.38 MeV above the $3\alpha$ threshold. The $0_2^+$ state was shown to be dominated by the channel, $I$ = $l$ = 0. Here $I$ and $l$ indicate the spin of $^8$Be and the orbital angular momentum of the relative motion between $^8$Be and alpha particle, respectively \cite{uegaki}. This is consistent with the result of the semi-microscopic calculation with OCM (orthogonality condition model) made by Horiuchi, where three alpha particles weekly interact in relative $S$-waves \cite{hori}. The $0_2^+$ state was shown to have a large reduced $\alpha$-width, $\theta^2 (a)$ = 0.68 in units of the Wigner limit, $\gamma_W^2$ = $3\hbar^2/2\mu a^2$, when the channel radius, $a$ = 6 fm, for $I=l=0$. Here $\mu$ is the reduced mass between $^8$Be and the $\alpha$ particle and $\theta^2(a) = \gamma^2 / \gamma_W^2$, where $\gamma ^2$ is the reduced $\alpha$-width (see Eq. (\ref{eq:28}) of section \ref{subsec:accc}). 

According to the suggestion made in Ref. \cite{thsr} mentioned above, we investigated the $0_2^+$ state of $^{12}$C by using a deformed alpha condensate wave function which only slightly deviates from the spherical one, Eq. (\ref{eq:1}) \cite{cbec}. We found that each of the $0_2^+$ wave functions, obtained by the full $3\alpha$ microscopic calculations of Refs. \cite{uegaki} and \cite{kamimura}, has a large squared overlap value of more than $90$ \% with the single condensate wave function of $3\alpha$ gas-like structure. This result gives decisive theoretical evidence that the $0_2^+$ state can be understood as a $3\alpha$ condensed state. 

The present paper will mainly focus on the exploration of the excitation modes of the finite alpha condensate by studying the states of $^{12}$C with quantum numbers $J^\pi=0^+$, $2^+$, and $4^+$ with excitation energies less than about 15 MeV. The exploration is made by using an $\alpha$ condensate wave function with spatial deformation. In spite of the fact that the existence of the $2_2^+$ state has been suggested at around 3.0 MeV above the $3\alpha$ threshold by the former theoretical works \cite{carbon,uegaki,kamimura}, it was only quite recently that the $2_2^+$ state was observed at $2.6\pm 0.3$ MeV above the $3\alpha$ threshold together with the alpha decay width, $1.0\pm 0.3$ MeV \cite{itoh}. Morinaga expected the existence of a rotational band with the $0_2^+$ state as the band head and assumed a $3\alpha$ linear chain structure as the intrinsic structure of the band \cite{mori}. However, as already mentioned above, this assumption is now hardly acceptable because the $3\alpha$ structure of the $0_2^+$ state is far from that of a linear chain configuration of $3\alpha$ particles. The $2_2^+$ state, in view of the microscopic $3\alpha$ cluster model calculations \cite{uegaki}, is dominated by the channel, $I=0$ and $l=2$, and has a large reduced $\alpha$-width amplitude, $\theta ^2 (a)=0.63$, with the channel radius, $a=6.0$ fm. This fact indicates not only that this state has a well developed $3\alpha$ cluster structure like the $0_2^+$ state, but also that the $2_2^+$ state will be able to be classified in the same family as the $0_2^+$ state in which the single channel, $I=l=0$ is dominant.
  Besides the $2_2^+$ state, the $0_3^+$ state was observed at 3.0 MeV above the $3\alpha$ threshold together with a broad width, $3.0\pm 0.7$ MeV \cite{ajze}. However, there are no calculations that reproduce the energy of this state satisfactorily. Though the binding energy is higher by about 3 MeV than experimental observation, the $0_3^+$ state obtained in Ref. \cite{uegaki} has large reduced $\alpha$-width amplitudes of both channels, [$I \otimes l$] = [$0 \otimes 0$] and [$2 \otimes 2$]. The density distribution of the $0_3^+$ state calculated by the AMD method (antisymmetrized molecular dynamics) appears to be somewhat similar to a linear chain configuration of three alpha particles \cite{en'yo}. Recently a very similar result was reported about the $0_3^+$ state by FMD (Fermionic molecular dynamics) calculation \cite{neff}.

In order to explore the excitation modes of the $3\alpha$ condensate, we study the $2_2^+$ and $0_3^+$ states by a GCM (generator coordinate method) calculation in which deformed $3\alpha$ condensate wave functions are superposed. This type of GCM calculation was successfully used in our previous paper \cite{cbec} for the study of the $0_2^+$ state. However the GCM calculation was not successful for the $2_2^+$ and $0_3^+$ states, because this GCM calculation is a bound state approximation in spite of the fact that these states are resonances with rather large widths of more than 1 MeV while the width of the $0_2^+$ state is very small of only a few eV. 
 Thus, in order to overcome this defect, we apply, beyond the bound state approximation, the method of ACCC (analytic continuation in the coupling constant) to our previous formalism. This method was first proposed by Kukulin {\it et al.} \cite{kuku}, and is now recognized as a powerful method to calculate the energy and total width of a resonance state \cite{tanaka, aoyama}. 

The main result of the present paper is that the microscopic $3\alpha$ condensate wave function with spatial deformation is able to reproduce consistently both energy and total width of the recently observed $2_2^+$ state by the application of the method of ACCC. Furthermore, the $2_2^+$ state is found to be mostly represented by a single $3\alpha$ condensate wave function with slightly deformed shape, where the resonance wave function of the $2_2^+$ state is obtained by the use of our newly devised technique which allows us to extract correct resonance wave functions even within the formalism of bound state approximation. The density distribution is shown to be almost the same as that of the $0_2^+$ state which is considered as a $3\alpha$ Bose-condensed state, if the binding energy of the $2_2^+$ state is adjusted to the same value as the one of the $0_2^+$ state. The kinetic energy, nuclear interaction energy, and Coulomb interaction energy of the calculated $2_2^+$ state are further shown to be very similar to those of the $0_2^+$ state. These results indicate that the $2_2^+$ state has a structure closely related to the $0_2^+$ state where three alpha particles form a gas-like structure with dilute density. On the other hand, we found that our present functional space of the $\alpha$ condensate wave function in which strongly prolate condensed wave functions are not included is, presumably, not well adapted to represent the $0_3^+$ state. This point of view may be reasonable if the $0_3^+$ state has $3\alpha$ linear-chain-like structure. 

The content of this paper is as follows. In section \ref{sec:form}, the formulation of the present aproach is explained. A brief description of our GCM approach in which the deformed condensate wave functions are superposed after angular momentum projection and the complete elimination of the center-of-mass spurious components is given. The ACCC method is also outlined there. The results of the GCM calculations for the $0_1^+$, $2_1^+$, $4_1^+$, and $0_2^+$ states are given in subsection \ref{subsec:gcm}. The states are treated within the bound state approximation. In subsection \ref{subsec:resaccc}, the calculations of the energies and widths of the $2_2^+$ and $4_2^+$ states are presented in which the method of ACCC is used. The wave functions of the $2_2^+$ and $4_2^+$ are analyzed in the following subsection \ref{subsec:analyses}. There we introduce a new approximate procedure to estimate the wave functions of the resonances. In subsection \ref{subsec:0304+}, we calculate the energies and widths of the $0_3^+$ and $0_4^+$ states by using the method of ACCC. We make the same analyses for the wave functions as in subsection \ref{subsec:analyses}. In subsection \ref{subsec:disc1}, we check how much $J^\pi$ $=$ $0^+$ and $2^+$ components are contained in the deformed intrinsic wave function of the alpha condensate. In subsection \ref{subsec:disc2}, we show that if only the orthogonality between the $0_1^+$ and $0_2^+$ states is imposed, the structure of the $0_2^+$ state is not influenced by the detailed structure of the $0_1^+$ state. In subsection \ref{subsec:analysis}, we demonstrate that the oblate and prolate deformed intrinsic wave functions of the alpha condensate give similar wave functions after the angular momentum projection. Conclusions are given in section \ref{sec:conc}. In the APPENDIX, we give the explicit calculus of the density distribution in which the center-of-mass spurious components are completely eliminated.

\section{Formulation}\label{sec:form}
\subsection{Deformed intrinsic $n\alpha$ condensed wave function}\label{ssec:wf}
The deformed $n\alpha$ condensed wave function was first proposed in Ref. \cite{fthsr} as a natural modification of the spherical shape of wave function, Eq. (\ref{eq:1}). In this subsection, we explain how to derive the angular momentum projected wave function from the deformed $n\alpha$ condensed wave function.
The deformed $n\alpha$ condensed state can be expressed as given below, taking $C^\dagger_\alpha$ as a creation operator of an $\alpha$ particle with tri-axial variational parameters, $\beta_x, \beta_y, \beta_z$,
\begin{equation}
  |\Phi_{n\alpha} \rangle = (C_\alpha^\dagger)^n |{\rm vac} \rangle ,  \label{eq:2}
\end{equation}
where 
\begin{eqnarray}
\lefteqn{ C_\alpha^\dagger = \int d^3 R\  
     \exp \Bigl( - \frac{R_x^2}{\beta_x^2} - \frac{R_y^2}{\beta_y^2} - 
     \frac{R_z^2}{\beta_z^2} \Bigr)
     \int d^3 r_1  \cdots d^3 r_4 }\hspace{8cm} \nonumber \\
     \times \varphi_{0s}({\vc r}_1 - {\vc R}) 
     a_{\sigma_1 \tau_1}^\dagger ({\vc r}_1) 
     \cdots \varphi_{0s}({\vc r}_4 - {\vc R}) 
     a_{\sigma_4 \tau_4}^\dagger ({\vc r}_4). \label{eq:3}
\end{eqnarray}
In the above equation, $a_{\sigma \tau}^\dagger({\vc r})$ is a creation operator of one nucleon with spin $\sigma$, isospin $\tau$ at a spatial point ${\vc r}$, and $\varphi_{0s}({\vc r}- {\vc R})$ is a $0s$ harmonic oscillator wave function around a center ${\vc R}$ with the size parameter $b$ as follows,
\begin{equation}
 \varphi_{0s}({\vc r}-{\vc R}) = (\pi b^2)^{-3/4} \exp\Big\{-\frac{({\vc r}-{\vc R})^2}{2 b^2}\Big\}. \label{eq:4}
\end{equation}
In order to write down the wave function of the deformed $n\alpha$ condensed state (\ref{eq:2}) in coordinate space and also to perform numerical calculations of matrix elements of physical quantities, it is necessary to introduce Brink's $n\alpha$ cluster model wave function \cite{brink}, $\Phi^{\rm B}({\vc R}_1, \cdots, {\vc R}_n)$ defined as
\begin{eqnarray}
\lefteqn{\Phi^{\rm B}({\vc R}_1, \cdots, {\vc R}_n)
 \equiv \frac{1}{\sqrt{(4n) !}} \det \{ \varphi_{0s}({\vc r}_1 - {\vc R}_1) 
     \chi_{\sigma_1 \tau_1}}\hspace{6cm} \nonumber \\ 
     \cdots \varphi_{0s}({\vc r}_{4n} - {\vc R}_n) \chi_{\sigma_{4n} \tau_{4n}} \}. \label{eq:5}
\end{eqnarray}
Here $\chi_{\sigma \tau}$ is a spin-isospin wave function of a nucleon. From Eqs. (\ref{eq:2}), (\ref{eq:3}), the deformed $n\alpha$ condensed wave function can be written as a superposition of the Brink's $n\alpha$ cluster model wave functions multiplied by a single deformed Gaussian wave packet of ${\vc R}_1,\cdots , {\vc R}_n$ with the width, $\beta_i, (i=x,y,z)$, in the following way,
\begin{eqnarray}
&& \Phi_{n\alpha}(\vc{\beta}) \equiv
 \langle \,{\vc r}_1 \sigma_1 \tau_1, \cdots {\vc r}_{4n} \sigma_{4n} 
 \tau_{4n} |\Phi_{n\alpha}\, \rangle \nonumber \\ 
&& = \int\! d^3\! R_1\cdots d^3\! R_n\ 
     \exp\! \Big\{ -\sum_{i=1}^n \bigl( \frac{R_{ix}^2}{\beta_x^2} + 
     \frac{R_{iy}^2}{\beta_y^2} + \frac{R_{iz}^2}{\beta_z^2}\bigr) \Big\} \nonumber \\ 
&& \hspace{3cm}\times \Phi^{\rm B}({\vc R}_1, \cdots, {\vc R}_n),\nonumber \\ 
&&\label{eq:6} \\
&& \vc{\beta} \equiv (\beta_x, \beta_y, \beta_z). \nonumber \\
&&\label{eq:6a}
\end{eqnarray}
It is easy to perform the above integration over the variables, ${\vc R}_1,\cdots , {\vc R}_n$, as shown in Ref. \cite{fthsr} in detail. Consequently the deformed $n\alpha$-condensed wave function is simply given by
%\begin{widetext}
\begin{equation}
 \Phi_{n\alpha}(\vc{\beta}) \propto {\cal A}\ \Big[ \prod_{i=1}^n\exp \Big\{-\Bigl(\frac{2 X_{ix}^2}{B_x^2}  + \frac{2 X_{iy}^2}{B_y^2} + \frac{2 X_{iz}^2}{B_z^2}\Bigr) \Big\} \phi(\alpha_i) \Big]. \label{eq:7}
\end{equation}
%\end{widetext}
Here, $B_k^2 = b^2 + 2\beta_k^2$, ($k=x,y,z$). The other notations are the same as used in Eq. (\ref{eq:1}). ${\cal A}$ is the antisymmetrizer operating on all nucleons in the system. It should be noted that Eq. (\ref{eq:7}) shows that the center-of-mass motions of $n\alpha$ particles occupy the same deformed orbit, $\exp (-2X_x^2 / B_x^2 -2 X_y^2 / B_y^2 -2 X_z^2 / B_z^2 )$. It is also to be noted that the normalized $n\alpha$ condensed wave function, Eq. (\ref{eq:7}) coincides with the shell model Slater determinant when $\beta_i \rightarrow 0 \ (i=x,y,z)$ as a limit. This is an important property that is taken over from Brink's wave function. On the contrary, the $n\alpha$ condensed wave function corresponds to a free $n\alpha$ particle state in which $n\alpha$ particles do not correlate with each other when $\beta_i \rightarrow \infty \ (i=x,y,z)$. 

The $\alpha$ condensed wave function with good quantum number of angular momentum is derived by projecting out the angular momentum from the deformed $\alpha$ condensed wave function in the following way,
%\begin{widetext}
\begin{eqnarray}
&& \Phi_{n\alpha}^J(\vc{\beta}) = \int d\Omega 
 D^{J*}_{M K}(\Omega) {\widehat R}(\Omega) \Phi_{n\alpha}(\vc{\beta}) \nonumber \\ 
 &&= \int d\Omega D^{J*}_{M K}(\Omega) \int d^3 R_1\ \cdots d^3 R_n \times \nonumber \\ 
 && \exp \Bigl( -\sum_{i=1}^n \sum_{k=x,y,z} \frac{R_{ik}^2}{\beta_k^2} \Bigr) \Phi^{\rm B}(R^{-1}(\Omega){\vc R}_1, \cdots,  R^{-1}(\Omega){\vc R}_n) \nonumber \\ 
 && = \int d\Omega D^{J*}_{M K}(\Omega) \int d^3 R_1\ \cdots d^3 R_n \times \nonumber \\ 
&& \exp \Big\{ -\sum_{i=1}^n \sum_{k=x,y,z} \frac{(R(\Omega){\vc R}_i)_k^2}
 {\beta_k^2} \Big\} \Phi^{\rm B}({\vc R}_1, \cdots, {\vc R}_n), \nonumber \\
&& \label{eq:8}
\end{eqnarray}
%\end{widetext}
where $\Omega$ is the Euler angle, $D^J_{MK}(\Omega)$ the Wigner $D$-function, $\widehat{R}(\Omega)$ the rotation operator, and $R(\Omega)$ the $3\times 3$ rotation matrix corresponding to $\widehat{R}(\Omega)$. Here use is made of the relation $\varphi_{0s}(R(\Omega){\vc r} - {\vc R}) = \varphi_{0s}({\vc r} - R^{-1}(\Omega){\vc R})$. We should note that the rotation with respect to the total angular momentum is equivalent to the rotation with respect to the total orbital angular momentum, since the intrinsic spins of the $\alpha$ clusters are saturated.

Throughout the present paper, all calculations are performed with the restricton to axially symmetric deformation in the same way as our previous work, Ref. \cite{fthsr}, where the $z$-axis is taken as the symmetry axis: $\beta_x = \beta_y \ne \beta_z$. Thus, the formulas used in Eq. (\ref{eq:8}) are simplified as follows, 
\begin{eqnarray}
 \left\{
  \begin{array}{lcl}
\int d\Omega         &\longrightarrow & 4\pi\int d\cos \theta \\ 
D^{J*}_{M K}(\Omega)  &\longrightarrow & d^{J}_{M 0}(\theta) \\ 
{\widehat R}(\Omega) &\longrightarrow & {\widehat R}_y(\theta) \\ 
R(\Omega)            &\longrightarrow & R_y(\theta). \label{eq:9}
  \end{array}
 \right.
\end{eqnarray}
In all the following sections, we make use of the notation, $\vc{\beta}$ to express $(\beta_x=\beta_y, \beta_z)$.

\subsection{Elimination of spurious center-of-mass motion}\label{subsec:cm}

The total center-of-mass motion can be easily separated out of the $n\alpha$ condensed wave function (\ref{eq:7}) as follows,
\begin{eqnarray}
 \Phi_{n\alpha}(\vc{\beta})\hspace{-0.15cm} &\propto& \hspace{-0.15cm} \exp \Big\{ -\frac{2n}
 {B_x^2} (X_{Gx}^2\! +\! X_{Gy}^2) - \frac{2n}{B_z^2} X_{Gz}^2 \Big\} {\widehat \Phi}_{n\alpha}(\vc{\beta}), \nonumber  \\
 & &  \label{eq:10a}\\
 {\widehat \Phi}_{n\alpha}(\vc{\beta})\hspace{-0.15cm} &=&\hspace{-0.15cm} {\cal A} \Big[ \prod_{i=1}^n\exp \Big\{\!\! -\!\! \sum_{k=x,y,z}\!\! \frac{2}{B_k^2} (X_{ik}\!\! -\!\! X_{Gk})^2 \Big\} \phi(\alpha_i) \Big],\nonumber \\ 
  & & \label{eq:10}
\end{eqnarray}
Clearly ${\widehat \Phi}_{n\alpha}(\vc{\beta})$ does not contain the total center-of-mass coordinate, ${\vc X}_G$, of the system and it is the internal wave function of $\Phi_{n\alpha}(\vc{\beta})$. All the calculations are not made with $\Phi_{n\alpha}(\vc{\beta})$ but with $\widehat{\Phi}_{n\alpha}(\vc{\beta})$ which is an eigenstate of total momentum with eigenvalue zero. Thus, the proper $\alpha$-condensed wave function with a good quantum number of angular momentum is not $\Phi^J_{n\alpha}(\vc{\beta})$ represented by Eq. (\ref{eq:8}) but ${\widehat \Phi}^J_{n\alpha}(\vc{\beta})$ defined as

\begin{equation}
 {\widehat \Phi}^J_{n\alpha}(\vc{\beta}) = \int d\cos 
 \theta \,d^J_{M0}(\theta) {\widehat R}_y(\theta) \ 
 {\widehat \Phi}_{n\alpha}(\vc{\beta}). \label{eq:11}
\end{equation}
In later subsections we use the notation ${\widehat \Phi}^{{\rm N},J}_{3\alpha}(\vc{\beta})$ in order to express the normalized state of ${\widehat \Phi}^{J}_{3\alpha}(\vc{\beta})$
\begin{equation}
{\widehat \Phi}_{3\alpha}^{{\rm N},J}(\vc{\beta})
=\frac{{\widehat \Phi}_{3\alpha}^J(\vc{\beta})}{\sqrt{\big\langle{\widehat \Phi}_{3\alpha}^J(\vc{\beta}) \big| {\widehat \Phi}_{3\alpha}^J(\vc{\beta})\big\rangle}}. \label{eq:25}
\end{equation}
Since in Ref. \cite{fthsr}, we have given a detailed explanation of how to derive the matrix elements of a general translationally invariant scalar operator $\widehat{O}$ using the wave function with good quantum number of angular momentum, ${\widehat \Phi}^J_{n\alpha}(\vc{\beta})$, we only give a brief outline here. While in Ref. \cite{fthsr}, we examined only the diagonal components of the matrix elements in the parameter space $\vc{\beta}$, we, in this paper, need to refer to the case of the off-diagonal components. We can write down the matrix elements in which the contribution of the spurious center-of-mass motion is eliminated, as follows,
\begin{widetext}
\begin{eqnarray}
&& \frac {\big\langle {\widehat \Phi}^J_{n\alpha}(\vc{\beta}) 
 \big|{\widehat O}\big|{\widehat \Phi}^J_{n\alpha}(\vc{\beta^\prime}) 
 \big\rangle}{\sqrt{\big\langle {\widehat \Phi}^J_{n\alpha}(\vc{\beta}) \big|{\widehat \Phi}^J_{n\alpha}(\vc{\beta}) \big\rangle \big\langle {\widehat \Phi}^J_{n\alpha}(\vc{\beta}^\prime) \big|{\widehat \Phi}^J_{n\alpha}(\vc{\beta^\prime}) \big\rangle}} \nonumber \\ 
 && = \frac{\int d\cos \theta \,d^J_{00}(\theta) \big\langle 
 {\widehat \Phi}_{n\alpha}(\vc{\beta})\big|{\widehat O}
 {\widehat R}_y(\theta)\big|{\widehat \Phi}_{n\alpha}(\vc{\beta^\prime}) \big\rangle}{\sqrt{\int d\cos \theta \,d^J_{00}(\theta) \big\langle 
 {\widehat \Phi}_{n\alpha}(\vc{\beta})|
 {\widehat R}_y(\theta)\big|{\widehat \Phi}_{n\alpha}(\vc{\beta}) \big\rangle \cdot \int d\cos \theta \,d^J_{00}(\theta) \big\langle 
 {\widehat \Phi}_{n\alpha}(\vc{\beta}^\prime)|
 {\widehat R}_y(\theta)\big|{\widehat \Phi}_{n\alpha}(\vc{\beta^\prime}) \big\rangle}} \nonumber \\
 && = \frac{\int d\cos \theta \,d^J_{00}(\theta)\big\langle 
 \Phi_{n\alpha}(\vc{\beta})\big|{\widehat O}
 {\widehat R}_y(\theta)\big|\Phi_{n\alpha}(\vc{\beta^\prime}) 
 \big\rangle /
  P_0(\theta)}{\sqrt{\int d\cos \theta \,d^J_{00}(\theta) \big\langle 
 \Phi_{n\alpha}(\vc{\beta})\big|{\widehat R}_y(\theta)
 \big|\Phi_{n\alpha}(\vc{\beta}) \big\rangle /P_1(\theta)\cdot \int d\cos \theta \,d^J_{00}(\theta) \big\langle
 \Phi_{n\alpha}(\vc{\beta}^\prime)\big|{\widehat R}_y(\theta)
 \big|\Phi_{n\alpha}(\vc{\beta^\prime}) \big\rangle /P_2(\theta)}}. \nonumber \\ 
 &&\label{eq:12}
\end{eqnarray}
\end{widetext}
Here the functions $P_0(\theta)$, $P_1(\theta)$, and $P_2(\theta)$ stem from the degrees of freedom of the center-of-mass motion and can be easily calculated analytically. We give the explicit form of $P_0(\theta)$ below. The expression of $P_1(\theta)$ is given by simply changing $B_x^\prime$, $B_y^\prime$, and $B_z^\prime$ that appear in the following expression of $P_0(\theta)$, to $B_x$, $B_y$, and $B_z$. And the expression of $P_2(\theta)$ is similarly given by simply changing $B_x$, $B_y$, and $B_z$ that appear in the following expression of $P_0(\theta)$, to $B_x^\prime$, $B_y^\prime$, and $B_z^\prime$,
\begin{widetext}
\begin{eqnarray}
P_0(\theta) &=& \Big\langle \exp \Bigl( - \sum_{k=x,y,z} 
 \frac{2n}{B_k^2} X_{Gk}^2 \Bigr) \Big| {\widehat R}_y(\theta)\Big| \exp \Bigl( 
 - \sum_{k=x,y,z} \frac{2n}{B_k^{\prime2}} X_{Gk}^2 \Bigr) \Big\rangle 
 \nonumber \\
 &=& \sqrt{ \Bigl(\frac{\pi}{2n}\Bigr)^3 \frac{B_x^2 B_x^{\prime2} B_y^2 B_y^{\prime2} B_z^2 B_z^{\prime2}}
 {(B_y^2+B_y^{\prime2})\{ (B_x^2+B_x^{\prime2})(B_z^2+B_z^{\prime2}) + (B_x^2 - B_z^2)(B_x^{\prime2} - B_z^{\prime2}) \sin^2 \theta \}} }. \label{eq:13}
\end{eqnarray}
\end{widetext}

\subsection{Approximate equivalence between prolate and oblate deformed wave functions after angular momentum projection with the exception of strongly prolate deformation}\label{subsec:analysis}

\begin{figure}[htbp]
\begin{center}
\includegraphics[scale=.85]{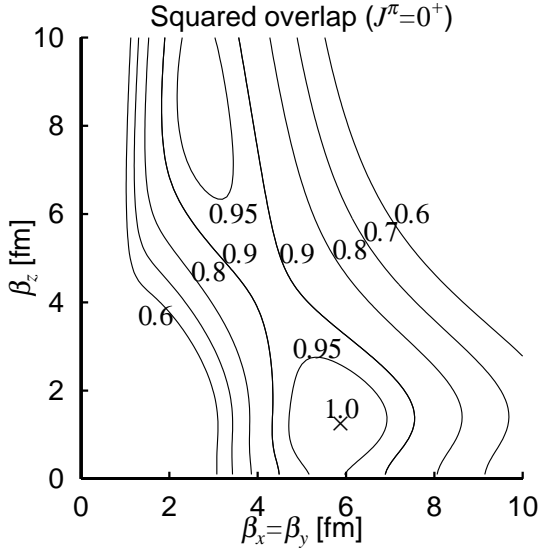}
\end{center}
\caption{Contour map of the squared overlap, $| \langle {\widehat \Phi}_{3\alpha}^{{\rm N},J=0}(\vc{\beta}) | {\widehat \Phi}_{3\alpha}^{{\rm N},J=0}(\vc{\beta_1}) \rangle |^2$, in the two-parameter space, $\beta_x(=\beta_y)$ and $\beta_z$, where $\vc{\beta}_1$$=$($\beta_x$$=$$\beta_y$$=$5.7 fm, $\beta_z$$=$$1.3$ fm). The harmonic oscillator size parameter $b$$=$$1.35$ fm.}\label{fig:analysis1}
\end{figure}

The $n\alpha$ condensate wave function with good angular momentum ${\widehat \Phi}^J_{n\alpha}(\vc{\beta})$ has a very characteristic property that ${\widehat \Phi}^J_{n\alpha}(\vc{\beta})$ obtained from a prolate intrinsic state can be obtained approximately from an oblate intrinsic state and vice versa, except for the case of strongly prolate deformation. In FIG. \ref{fig:analysis1}, we show the squared overlap, $| \langle {\widehat \Phi}_{3\alpha}^{{\rm N},J=0}(\vc{\beta}) | {\widehat \Phi}_{3\alpha}^{{\rm N},J=0}(\vc{\beta_1}) \rangle |^2$, in the parameter space, $\vc{\beta}$, where $\vc{\beta_1} \equiv (5.7\ {\rm fm}, 1.3\ {\rm fm})$. FIG. \ref{fig:analysis1} shows that the $0^+$ wave functions with prolate and oblate deformation are nearly equivalent to each other. On the contrary to FIG. \ref{fig:analysis1}, we show in FIG. \ref{fig:analysis10} the squared overlap $|\langle {\widehat \Phi}^{{\rm N}, J=0}_{3\alpha}(\vc{\beta})| {\widehat \Phi}^{{\rm N}, J=0}_{3\alpha}(\vc{\beta}_L)\rangle |^2$ in the parameter space of $\vc{\beta}$, where $\vc{\beta}_L$$=$$(0.1$ fm, $4.0$ fm) which expresses a strongly prolate deformation. This figure shows that the angular-momentum-projected wave function obtained from a strongly prolate deformation with $\beta_x$$=$$\beta_y$$<$0.5 fm can never be similar to any projected wave function from oblate deformation. 

We show below that the seemingly strange behavior of FIG. \ref{fig:analysis1} is a generic results in $n\alpha$ systems.

At first, we rewrite Eq. (\ref{eq:10}) in terms of the Jacobi coordinates,
\begin{equation}
 {\widehat \Phi}_{n\alpha}(\vc{\beta})\hspace{-0.15cm} =\hspace{-0.15cm} {\cal A} \Big[ \exp \Big\{\!\! - \! 2\! \sum_{i=1}^{n-1} \mu_i \Bigl(\frac{\xi_{ix}^2+\xi_{iy}^2}{B_x^2}\!+\!\frac{\xi_{iz}^2}{B_z^2}\Bigr) \Big\} \phi^n(\alpha) \Big], \label{eq:35}
\end{equation}
where the Jacobi coordinates, $\vc{\xi} _i$ and ${\vc X}_G$ $(i=1,\cdots, n-1)$, satisfy the following relation, 
\begin{eqnarray}
\left\{
 \begin{array}{c c l}
{\vc \xi}_i &=& {\vc X}_{i+1}-\frac{1}{i}\sum_{j=1}^i {\vc X}_j  \\
 {\vc X}_G   &=& \frac{1}{n}\sum_{j=1}^n {\vc X}_j \nonumber \\
      \mu_i  &=& \frac{i}{i+1} \hspace{1cm} (i=1,\cdots, n-1). \label{eq:36}
 \end{array}
 \right.
\end{eqnarray}
Furthermore, in Eq. (\ref{eq:35}), the exponent, $(\xi_{ix}^2+\xi_{iy}^2)/B_x^2+\xi_{iz}^2/B_z^2$, can be rewritten as
\begin{eqnarray*}
&&\frac{\xi_{ix}^2+\xi_{iy}^2}{B_x^2}+\frac{\xi_{iz}^2}{B_z^2} = \Bigl( \frac{2}{3B_x^2}+\frac{1}{3B_z^2} \Bigr) {\vc \xi}_i^2 + \Bigl( \frac{1}{3B_z^2}-\frac{1}{3B_x^2} \Bigr) \nonumber \\ 
 &&\hspace{3cm} \times (2\xi_{iz}^2-\xi_{ix}^2-\xi_{iy}^2) \nonumber \\
 && = \Bigl( \frac{2}{3B_x^2}+\frac{1}{3B_z^2} \Bigr) {\vc \xi}_i^2 + \Bigl( \frac{1}{3B_z^2}-\frac{1}{3B_x^2} \Bigr) \sqrt{\frac{16\pi}{5}}{\vc \xi}_i^2 Y_{20}(\vc{\widehat{\xi}}_i) ,\label{eq:37}
\end{eqnarray*}
where $\vc{\widehat{\xi}}_i$ is the polar angle of $\vc{\xi}_i$. Thus, the exponential term in Eq. (\ref{eq:35}), i.e. the relative coordinate parts of the wave function between $n$ alpha clusters, can be expanded as follows:

\begin{figure}[htbp]
\begin{center}
\includegraphics[scale=.85]{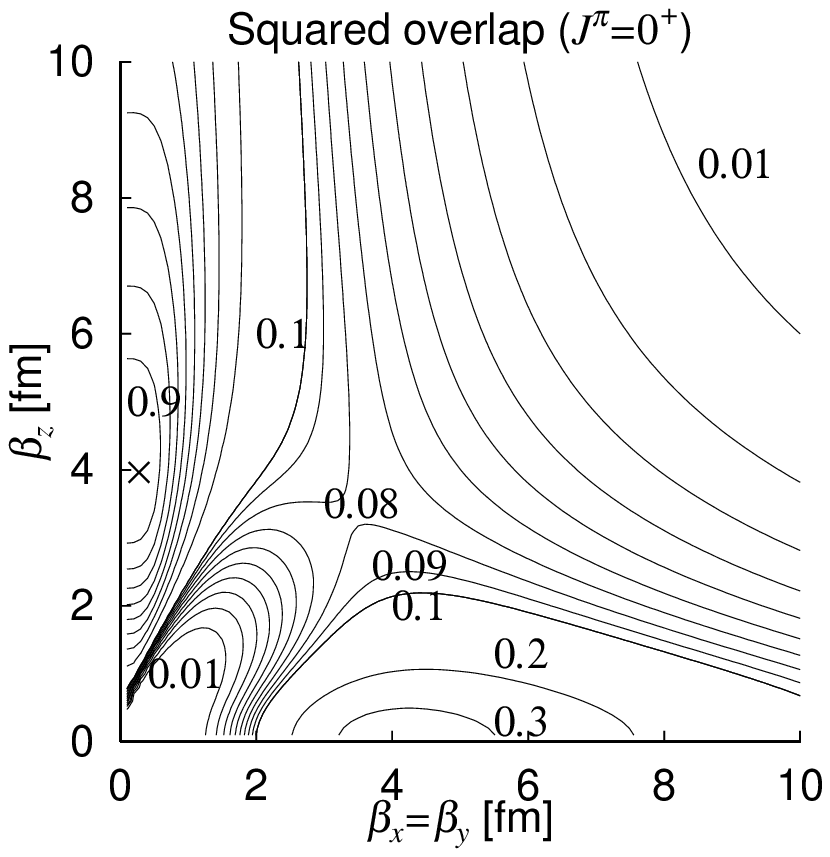}
\end{center}
\caption{Contour map of the squared overlap, $| \langle {\widehat \Phi}_{3\alpha}^{{\rm N},J=0}(\vc{\beta}) | {\widehat \Phi}_{3\alpha}^{{\rm N},J=0}(\vc{\beta}_L) \rangle |^2$, in the two-parameter space, $\beta_x(=\beta_y)$ and $\beta_z$, where $\vc{\beta}_1$$=$($\beta_x$$=$$\beta_y$$=$0.1 fm, $\beta_z$$=$$4.0$ fm). The harmonic oscillator size parameter $b$$=$$1.35$ fm.}\label{fig:analysis10}
\end{figure}

\begin{widetext}
\begin{eqnarray}
&& \exp \Big\{ -2 \sum_{i=1}^{n-1} \mu_i \Bigl(\frac{\xi_{ix}^2+\xi_{iy}^2}{B_x^2}+\frac{\xi_{iz}^2}{B_z^2}\Bigr) \Big\} = \exp \Big[ -2 \sum_{i=1}^{n-1} \mu_i \Big\{ \Bigl( \frac{2}{3B_x^2}+\frac{1}{3B_z^2} \Bigr) {\vc \xi}_i^2 + \Bigl( \frac{1}{3B_z^2}-\frac{1}{3B_x^2} \Bigr)\sqrt{\frac{16\pi}{5}}{\vc \xi}_i^2 Y_{20}(\vc{\widehat{\xi}}_i)\Big\} \Big] \nonumber \\ 
&&\hspace{2cm} =\exp \Big\{ -2\Bigl( \frac{2}{3B_x^2}+\frac{1}{3B_z^2} \Bigr) \sum_{i=1}^{n-1} \mu_i {\vc \xi}_i^2 \Big\} \Biggl[1 -2\Bigl( \frac{1}{3B_z^2}-\frac{1}{3B_x^2} \Bigr)\sqrt{\frac{16\pi}{5}} \sum_{i=1}^{n-1} \mu_i {\vc \xi}_i^2 Y_{20}(\vc{\widehat{\xi}}_i)+\cdots \Biggr] . \label{eq:38}
\end{eqnarray}
\end{widetext}
In the above expansion, the parameter, $(3B_z^2)^{-1}-(3B_x^2)^{-1}$, takes the opposite sign between prolate and oblate deformation so that the overlap between the wave functions, Eq. (\ref{eq:35}) with prolate and oblate deformation, cannot become large. However, once the wave function with good angular momentum is projected out of the wave function of Eq. (\ref{eq:35}), the above statement is no longer true. From the expansion of Eq. (\ref{eq:38}), the angular momentum projected wave function of the $0^+$ state takes the following form for the leading term,
\begin{equation}
{\cal A} \Big[ \exp \Big\{ -2\Bigl( \frac{2}{3B_x^2}+\frac{1}{3B_z^2} \Bigr) \sum_{i=1}^{n-1} \mu_i {\vc \xi}_i^2 \Big\} \phi^n(\alpha) \Big]. \label{eq:39}
\end{equation}
Also for the $2^+$ state, the leading term is 
%\begin{widetext}
\begin{eqnarray}
&&{\cal A} \Big[ \exp \Big\{ -2\Bigl( \frac{2}{3B_x^2}+\frac{1}{3B_z^2} \Bigr) \sum_{i=1}^{n-1} \mu_i {\vc \xi}_i^2 \Big\} \nonumber \\  
&& \hspace{2cm} \times \sum_{i=1}^{n-1} \mu_i\, {\vc \xi}_i^2\, Y_{20}(\vc{\widehat{\xi}}_i)\, \phi^n(\alpha) \Big]. \label{eq:40}
\end{eqnarray}
%\end{widetext}
Eqs. (\ref{eq:39}) and (\ref{eq:40}) show that the angular momentum-projected wave function of the $0^+$ state as well as the $2^+$ state can be parametrized by the sole coefficient $2(3B_x^2)^{-1}+(3B_z^2)^{-1}$. Thus, it follows that the angular momentum-projected wave functions from prolate and oblate intrinsic states cannnot become so much different from one another but rather they are close to each other as long as the values of $2(3B_x^2)^{-1}+(3B_z^2)^{-1}$ stay similar. It is to be noted that this analysis is true for a general $n\alpha$ system in as much as the leading terms of Eqs. (\ref{eq:39}), (\ref{eq:40}) are dominant.

\subsection{Hill-Wheeler equation}\label{hill}

In this subsection, we give a brief outline of the GCM approach. In order to discuss the quasi bound states as well as bound states, we solved the following Hill-Wheeler equation by superposing the wave functions of Eq. (\ref{eq:11}) with good quantum numbers of angular momentum in which the center-of-mass spurious components are completely eliminated,
\begin{equation}
\sum_{\vc{\beta^\prime}}
 \big\langle {\widehat \Phi}^J_{3\alpha}(\vc{\beta}) \big| (H-E) \big| {\widehat \Phi}^J_{3\alpha}(\vc{\beta^\prime}) \big\rangle f^J_\lambda (\vc{\beta^\prime}) =0 .  \label{eq:22}
\end{equation}
The solution of the above equation leads to the wave function as
\begin{equation}
\Psi^J_\lambda = \sum_{\vc{\beta}} f^J_\lambda (\vc{\beta}) {\widehat \Phi}^J_{3\alpha}(\vc{\beta}). \label{eq:23}
\end{equation}

In this paper, we do not superpose prolate deformed wave functions but only ${\widehat \Phi}^{J}_{3\alpha}(\vc{\beta})$ whose $\vc{\beta}$'s satisfy the relation $0.5$ fm $\leq$$\beta_z \leq \beta_x=\beta_y \leq 20.25$ fm. The numbers of the adopted wave functions are 61 and 75 for $J^\pi=0^+$ states and the other $J^\pi$ states, respectively. The reason of this choice of our functional space is that, as discussed in section \ref{subsec:analysis}, the prolate and oblate deformed wave functions are similar except for the case of strongly prolate deformation. The superposition of the similar wave functions sometimes causes numerical inaccuracy. We can confirm that the restricted functional space which we adopt is sufficient to reproduce the $0_1^+$, $2_1^+$, and $0_2^+$ states. For these three states, we solved Eq. (\ref{eq:22}) in our previous paper of Ref. \cite{cbec}, where both the oblate and prolate wave functions including strongly prolate ones are superposed. We will see that the binding energies and r.m.s. radii of these states obtained in the previous paper are completely the same as the present results given in TABLE \ref{tab:1} of section \ref{subsec:gcm}. 

For the $2_2^+$ state, we have found that the density distribution and the transition density between the $2_2^+$ and $0_1^+$ states obtained in this paper are almost equivalent to those calculated by the wave functions obtained by solving microscopic full-three-$\alpha$ problems without any model assumptions \cite{kamimura}. We will give the detailed discussion about this finding in our next paper. 

\subsection{Analytic continuation in the coupling constant}\label{subsec:accc}
Besides the $0^+_2$ state which has a narrow width, other excited states located above the $3\alpha$ threshold have non-negligible widths. In order to discuss resonance states that have broad widths, we go beyond the framework of bound state approximation by using the ACCC method.

In the ACCC method, we introduce an attractive pseudo potential, $V$, which is added to the microscopic Hamiltonian $H$ in the following way:
\begin{equation}
H^\prime(\delta) = H + \delta \times V ,\label{eq:14}
\end{equation}
where $\delta$ is a newly introduced coupling constant. The basic idea of the ACCC method is to distinguish resonance states from continuum states by increasing the coupling constant $\delta$ from the physical value, $\delta=0$, and making the interaction unphysically attractive. As a result of this operation, the energy of a resonance state goes below the $3\alpha$ threshold, while a continuum state never goes under the $3\alpha$ threshold. This operation allows us to handle the resonance states within the framework of the bound state approximation.

In this work, we adopted the following form for the pseudo potential,
\begin{equation}
V = -\frac{1}{2}\sum_{i\neq j}^{12} 80[{\rm MeV}] \exp \Big( - \frac{r_{ij}^2}{ (2.5[{\rm fm}])^2} \Big) .\label{eq:15}
\end{equation}
The reason why we assign 2.5 fm for the range parameter of the pseudo potential is that in order to pull down the resonance state efficiently into the bound state region, it is useful to make the range parameter of the pseudo potential larger than the range of the attractive part of the nuclear force. 

In order to treat the resonance states which are converted to bound states, we solve the following Hill-Wheeler equation for the ACCC Hamiltonian $H^\prime(\delta)$,

\begin{equation}
\sum_{\vc{\beta^\prime}}
 \left\langle {\widehat \Phi}^J_{3\alpha}(\vc{\beta}) \Big| \Big(H^\prime(\delta)-E^J_\lambda(\delta)\Big) \Big| {\widehat \Phi}^J_{3\alpha}(\vc{\beta^\prime}) \right\rangle  f^J_\lambda (\vc{\beta^\prime}, \delta) =0 .  \label{eq:27}
\end{equation}
The solution of the above equation is given as a function of the coupling constant $\delta$ as follows,
\begin{equation}
\Psi ^J_\lambda (\delta) = \sum_{\vc \beta}
 f^J_\lambda (\vc{\beta}, \delta) {\widehat \Phi}^J_{3\alpha}(\vc{\beta}).  \label{eq:30}
\end{equation}

Once the resonance state is obtained in the bound state region, the energy as a function of $\delta$ of the state in question can be approximated by a Pad\'e approximant that has the following form, 
\begin{eqnarray}
&& k_{[N,M]} = i \frac{c_0+c_1 x + c_2 x^2 + \dots + c_N x^N}{1+d_1 x + d_2 x^2 + \dots + d_M x^M} , \label{eq:16} \\ 
&& E(\delta)-E_{3\alpha}^{\rm th}(\delta) = k_{[N,M]}^2 .\label{eq:17}
\end{eqnarray}
Here $x=\sqrt{\delta-\delta _0}$, and the constant $\delta_0$ satisfies the relation, $E(\delta_0) = E^{\rm th}_{3\alpha}(\delta_0)=3E_\alpha(\delta_0)$. $E_\alpha(\delta)$ is the binding energy of an alpha particle as a function of $\delta$. We can determine the coefficients of the Pad\'e approximant of the right hand side in Eq. (\ref{eq:16}) using the values of the binding energy derived from the Hill-Wheeler equation, Eq. (\ref{eq:27}). While the energy function, $k^2[N,M]$, determined with the help of the Pad\'e approximation only has a real part in the bound state region below the $3\alpha$ threshold, it takes complex numbers above the $3\alpha$ threshold, where the binding energy and the total width will be given as ${\rm Re} (E(\delta)-E^{\rm th}_{3\alpha}(\delta))$ and $-2\cdot {\rm Im} (E(\delta)-E^{\rm th}_{3\alpha}(\delta))$, respectively. We can extrapolate the energy function in $\delta$ back to the physical point, $\delta = 0$, to reach the desired binding energy and width.

 The numerical accuracy of the binding energy which we obtain by the GCM calculation is approximately seven significant digits, namely the precision of several tens of eV since the binding energy is around 80 MeV. The determination of the coefficients $c_i$ and $d_i$ of the Pad\'e approximant is made by using $E(\delta)-E_{3\alpha}^{\rm th}$ which is of the order of $10^{-1}$ MeV $\sim 1$ MeV. Therefore, due to this subtraction of $E(\delta)-E_{3\alpha}^{\rm th}$, the numerical accuracy of the coefficients $c_i$ and $d_i$ is lower than that of the GCM binding energy and it atteins approximately four significant digits. If the width of a resonance is much smaller than the energy of the resonance by more than four orders of magnitude as this is the case for the $0_2^+$ state whose energy and width are about 0.4 MeV and about 9$\times$$10^{-6}$ MeV, respectively, we cannot calculate the width of the resonance with the ACCC method because of lack of numerical accuracy.

However, we can devise an approximate method to calculate such small widths within the framework of the ACCC method by using a region of negative values of $\delta$ where we can safely calculate the resonance width $\Gamma(\delta)$. With the condition that the width be calculated with good numerical accuracy, we choose a negative $\delta$, $\delta$$=$$\delta_z$$ < $$ 0$, whose magnitude $|\delta_z|$ is as small as possible. According to the $R$-matrix theory \cite{lane}, the decay width $\Gamma(\delta)$ is expressed as
\begin{eqnarray}
&&\Gamma(\delta) =2P_l (\epsilon(\delta))\cdot \gamma^2(\delta), \nonumber \\
&&P_l (\epsilon)=\frac{\rho}{F_l^2(\rho)+G_l^2(\rho)},\hspace{0.5cm} \rho=a\sqrt{\frac{2\mu}{\hbar^2}\epsilon},\label{eq:28}
\end{eqnarray}
where $P_l$ is the penetrability of the $^8$Be$(0^+)$$+$$\alpha$ decay channel with partial wave $l$, $\mu$ being the reduced mass of $^8$Be and $\alpha$, and $a$ the channel radius. Here $\epsilon(\delta)$ is the decay energy, $\epsilon(\delta)$$=$$E(\delta)$$-$$E^{J=0}_{2\alpha}(\delta)$, where $E^{J=0}_{2\alpha}(\delta)$ is the binding energy of $^8$Be$(0^+)$, and $F_l$ and $G_l$ are regular and irregular Coulomb functions, respectively. From this formula we obtain
\begin{equation}
\Gamma(\delta=0) \approx \frac{P_l(\bar{\epsilon})}{P_l(\epsilon(\delta_z))}\Gamma(\delta_z) \label{eq:penetrability}
\end{equation}
under the approximation $\gamma^2(\delta=0)$$\approx$$\gamma^2(\delta_z)$. Here, $\bar{\epsilon}$ is the observed decay energy into the $^8$Be$(0^+)$$+$$\alpha$ decay channel.

\subsection{Microscopic Hamiltonian}\label{subsec:force}
The microscopic Hamiltonian is expressed as
\begin{equation}  
 H = T - T_G + V_N + V_C. \label{eq:18}
\end{equation}
Here $T$ denotes the kinetic energy, $T_G$ the total center-of-mass kinetic energy, and $V_C$ the Coulomb interaction between protons. These quantities are given by 
\begin{eqnarray}
 T   &=& \sum_{i=1}^{12} \frac{-\hbar^2}{2 m} \Bigl( 
 \frac{\partial}{\partial {\vc r}_i} \Bigr)^2, \nonumber \\
 T_G &=& \frac{-\hbar^2}{24 m} \Bigl( \frac{\partial}{\partial {\vc X}_G} \Bigr)^2, \nonumber \\
 V_C &=& \frac{1}{2}\sum_{i\neq j}^{12} \frac{1-\tau_{zi}}{2} 
\frac{1-\tau_{zj}}{2} \frac{e^2}{r_{ij}}. \label{eq:19}
\end{eqnarray}
$V_N$ denotes the effective two-body nuclear interaction. In this paper, the Volkov No. 1 force with the value of Majorana parameter $M=0.575$ \cite{volkov} used by Uegaki {\it et al.} and the Volkov No.2 force with $M=0.59$ \cite{volkov} used by Kamimura {\it et al.} are adopted, which are expressed in the following formula: 
\begin{equation}
V_N = \frac{1}{2}\sum_{i\neq j}^{12}
\left\{(1-M)-MP_\sigma P_\tau\right\}_{ij} \sum_{n=1}^2v_n 
\exp\left(\!\!-\frac{r_{ij}^2}{a_n^2}\right). \label{eq:20}
\end{equation}

\section{Results}\label{sec:res}

\subsection{GCM study of the ground band states and the $0_2^+$ state}\label{subsec:gcm}
In our previous paper of Ref. \cite{cbec}, we concluded that the $0_2^+$ state of $^{12}$C has Bose-condensate character where the three alpha particles occupy the same center of mass $S$-state. Here we give some more detailed analyses to reinforce this conclusion.

In Ref. \cite{cbec}, we analysed the microscopic $3\alpha$ calculations by Uegaki {\it et al.} and Kamimura {\it et al.} which were performed about a quarter century ago. We then concluded that the $0_2^+$ wave functions obtained by them are almost the same as the $3\alpha$ Bose condensed wave functions. We here briefly recapitulate the essence of our argument in Ref. \cite{cbec}.

In Ref. \cite{uegaki}, Uegaki {\it et al.} adopted the Volkov No. 1 force with a Majorana parameter, $M=0.575$, as effective nuclear force, while in Ref. \cite{kamimura}, Kamimura and co-workers adopted Volkov No. 2 force with the value, $M=0.59$. Uegaki. {\it et al.} and Kamimura and co-workers followed different approaches, but in essence there was no difference between them since both approaches fully took into account the degrees of freedom of relative motions between the constituent alpha particles as well as the Pauli principle. In both approaches, the alpha particle with $(0s)^4$ of harmonic oscillator type was used with a fixed oscillator size parameter value, $b=1.41$ fm for Volkov No. 1 and $b=1.35$ fm for Volkov No. 2. For example, Kamimura and co-workers adopted the resonating group method (RGM) where the following equation was solved,
\begin{equation}
\big\langle \phi^3(\alpha)\big|(H-E)\big| {\cal A} \{\chi(\vc{s},\vc{t})\phi^3(\alpha)\} \big\rangle =0 . \label{eq:21}
\end{equation}
Here $\vc{s}$ and $\vc{t}$ are the Jacobi coordinates of the center-of-mass motion of three alpha clusters. We can clearly see that our model wave function, Eq. (\ref{eq:10}), mathematically corresponds to a specific form of the above complete three-body wave function ${\cal A} \{\chi(\vc{s},\vc{t})\phi^3(\alpha)\}$.

\begin{table*}
\begin{ruledtabular}
\begin{center}
\caption{Comparison of the GCM calculation solving the Hill-Wheeler equation, Eq. (\ref{eq:22}), with the full $3\alpha$ calculation \cite{uegaki, kamimura} with respect to the binding energy, $E$, and r.m.s radius, $R_{\rm r.m.s}$. The comparison is made for the $0^+_1$, $2^+_1$, $4_1^+$, and $0^+_2$ states for two cases of the effective two-nucleon force. The r.m.s radii of Ref. \cite{uegaki} which was cited in our previous paper \cite{cbec} were the charge radii containing the proton size $\sqrt{\langle r^2 \rangle _p}=0.813$ fm. Here we cite the values without proton size effect, which can be directly compared with our r.m.s radii. Note that for the $4_1^+$ state, the spurious components of continuum states are subtracted by the method discussed in section \ref{subsec:analyses}.}
\begin{tabular}{ c c c c c c c c c} 
\label{tab:1}
% \hline
 & \multicolumn{4}{c} {{\rm Volkov No.1} $M=0.575,\ E^{\rm th}_{3\alpha}=-81.01$\ MeV}
 & \multicolumn{4}{c} {{\rm Volkov No.2} $M=0.59 ,\ E^{\rm th}_{3\alpha}=-82.04$\ MeV} \\
% & \multicolumn{3}{ c }{$E^{\rm th}_{3\alpha}=-81.01$} 
% & \multicolumn{3}{ c }{$E^{\rm th}_{3\alpha}=-82.04$} \\
 \hline
 & \multicolumn{2}{c} {GCM calculation} & \multicolumn{2}{c} {full\ $3\alpha$\ calculation\ \cite{uegaki}} & \multicolumn{2}{c} {GCM calculation} &\multicolumn{2}{c} {full\ $3\alpha$\ calculation\ \cite{kamimura}}        \\
 \hline
$J^\pi$ & $E$ (MeV) &$R_{\rm r.m.s}$ (fm)&$E$ (MeV) &$R_{\rm r.m.s}$ (fm)
 & $E$ (MeV) &$R_{\rm r.m.s}$ (fm)&$E$ (MeV) &$R_{\rm r.m.s}$ (fm)\\
 \hline
 $0^+_1$ &$-87.81$&2.40&$-87.9$&2.40&$-89.52$&2.40&$-89.4$&2.40 \\
% \hline
 $2^+_1$ &$-85.34$&2.38&$-85.7$&2.36&$-86.71$&2.38&$-86.7$&2.38 \\
 $4_1^+$ &$-79.87$&2.30&$-80.4$&2.29&$-80.30$&2.31&$-80.0$ &2.31 \\
% \hline
 $0^+_2$ &$-79.97$&4.44&$-79.3$&3.40&$-81.79$&3.83&$-81.7$&3.47
% \hline
\end{tabular}
\end{center}
\end{ruledtabular}
\end{table*}

\begin{figure}[htbp]
\begin{center}
\includegraphics[width=14.5cm]{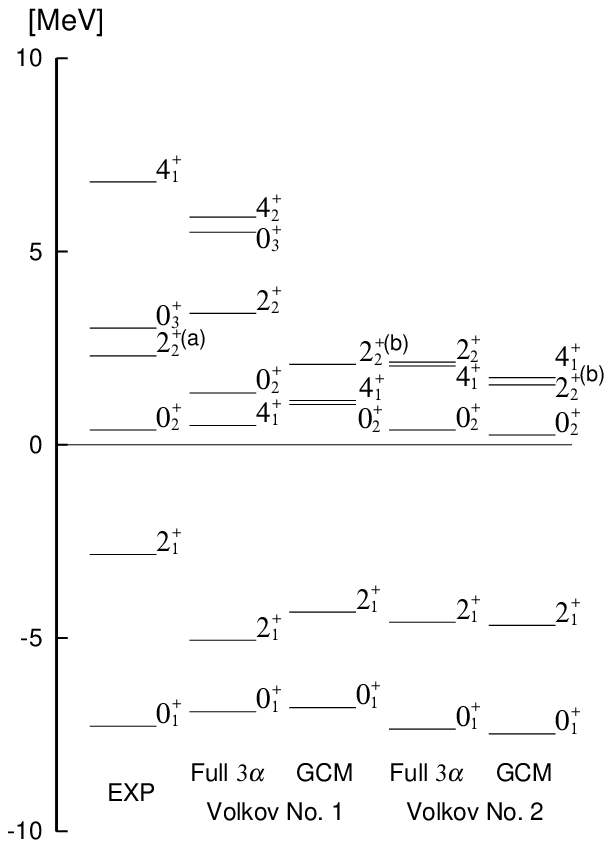}
\caption{Energy spectra of low lying positive parity states of $^{12}$C. ``Full $3\alpha$'' for the Volkov force, No. 1 and No. 2, are taken from \cite{uegaki} and \cite{kamimura}, respectively. \\ 
(a): the $2_2^+$ state in experiment comes from Ref. \cite{itoh}. The other observed states are taken from Ref. \cite{ajze}. \\
(b): GCM $+$ ACCC }
\label{fig:levels}
\end{center}
\end{figure}

Adopting the same nuclear forces as Uegaki {\it et al.} and Kamimura {\it et al.}, we solved the Hill-Wheeler equation of Eq. (\ref{eq:22}) for the ground state band with $0_1^+$, $2_1^+$, $4_1^+$, and the $0_2^+$ state. The eigen energies and r.m.s radii of our GCM calculations are compared with those by Uegaki {\it et al.} and Kamimura {\it et al.}, respectively, in TABLE \ref{tab:1}. The comparison of the eigen energies is also shown in FIG. \ref{fig:levels}. Throughout the $0_1^+, 2_1^+, 4_1^+$, and $0_2^+$ states, as well as our binding energies are almost the same as those obtained by Uegaki {\it et al.}, and also by Kamimura and co-workers. Considering the mini-max theorem of the variational problem and the fact that the complete $3\alpha$ model space contains our model space, we can say that our wave functions of those states are almost identical with the ones given by Uegaki {\it et al.} and by Kamimura and co-workers for each case of the adopted nuclear force. 

The calculated large root mean square radius of the $0_2^+$ state is quite remarkable. It is 4.44 fm for Volkov No. 1 and 3.83 fm for Volkov No. 2. The root mean square radii of the three states of the ground band are very similar and show that these states have a compact structure with normal density and do not have a pronounced three-alpha cluster structure. 

The most important result in Ref. \cite{cbec} is the fact that the $0_2^+$ wave function calculated in our GCM approach, therefore the $0_2^+$ wave function given by Uegaki {\it et al.} or by Kamimura {\it et al.}, is almost completely equivalent to the much simpler condensate wave function. In getting this result, the following steps were taken. At first we calculate the energy surface after the angular momentum projection to $J^\pi=0^+$ and obtained the minimum energy point at $\vc{\beta}_0 \equiv (1.5\ {\rm fm}, 1.5\ {\rm fm})$ where the minimum energy is calculated as $-87.68$ MeV for the case of Volkov No. 2 force. The minimum energy state is considered to correspond to the ground state.
We then calculated the projection operator,
\begin{equation}
P^{J=0}_\bot(\vc{\beta})=1-\big|{\widehat \Phi}_{3\alpha}^{N,J=0}(\vc{\beta}) \big\rangle \big\langle {\widehat \Phi}_{3\alpha}^{N,J=0}(\vc{\beta})\big|, \label{eq:261}
\end{equation}
at $\vc{\beta}=\vc{\beta_0}$. Here ${\widehat \Phi}_{3\alpha}^{{\rm N},J}(\vc{\beta})$ is the normalized state of ${\widehat \Phi}_{3\alpha}^{J}(\vc{\beta})$ as defined in Eq. (\ref{eq:25}).
The operator $P^{J=0}_\bot(\vc{\beta_0})$ creates the functional space orthogonal to the approximate ground state ${\widehat \Phi}^{J=0}_{3\alpha}(\vc{\beta_0})$. Next we calculated the energy surface within this functional space spanned by $P^{J=0}_\bot(\vc{\beta}_0)\widehat{\Phi}_{n\alpha}^{J=0}(\vc{\beta})$. The minimum energy point of this energy surface was obtained at $\vc{\beta}=\vc{\beta_1}\equiv (5.7\ {\rm fm}, 1.3\ {\rm fm})$ in the case of Volkov No. 2 force. Finally we calculated the squared overlap $|\langle \Psi_{\lambda=2}^{J=0} |C_1 P^{J=0}_\bot(\vc{\beta}_0)\widehat{\Phi}_{n\alpha}^{J=0}(\vc{\beta_1}) \rangle |^2$ with $C_1$ standing for the normalization constant of $P^{J=0}_\bot(\vc{\beta}_0)\widehat{\Phi}_{n\alpha}^{J=0}(\vc{\beta_1})$ and we found that this quantity is as large as 0.97. The fact that the $0_2^+$ state is represented by a simple state $P^{J=0}_\bot(\vc{\beta}_0)\widehat{\Phi}_{n\alpha}^{J=0}(\vc{\beta_1})$ naturally leads to the correctness of our picture that the $0_2^+$ state is considered to be a Bose condensed state composed of $3\alpha$ particles.

\begin{figure}[htbp]
\begin{center}
\includegraphics[scale=0.9]{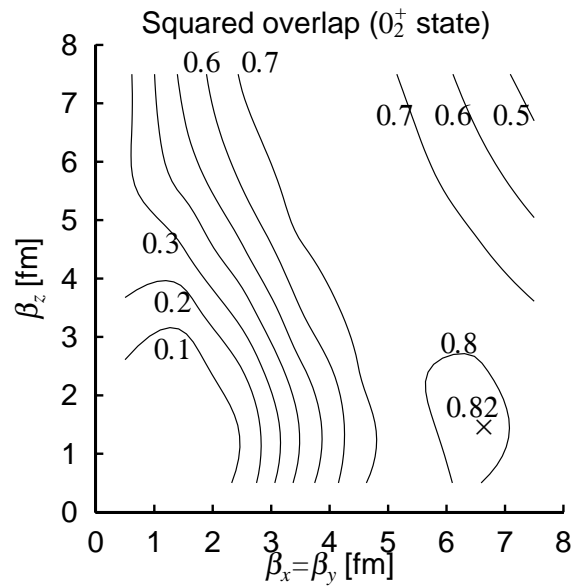}
\caption{Contour map of the squared overlap,\\ $| \langle {\widehat \Phi}_{3\alpha}^{{\rm N},J=0}(\vc{\beta}) | \Psi _{\lambda=2}^{J=0} \rangle |^2$, in the two-parameter space, $\beta_x(=\beta_y)$ and $\beta_z$. Numbers attached to the contour lines are the squared overlap values. The adopted effective nucleon force is Volkov No. 2.}
\label{fig:2}
\end{center}
\end{figure}

On the other hand, we here analyse the wave function of the $0_2^+$ state, $\Psi _{\lambda=2}^{J=0}$ by simply calculating the squared overlap, $| \langle {\widehat \Phi}_{3\alpha}^{{\rm N},J=0}(\vc{\beta}) | \Psi _{\lambda=2}^{J=0} \rangle |^2$, in the parameter space, $\vc{\beta}$. We show the contour map in FIG. \ref{fig:2}. These quantities indicate how much the $0_2^+$ wave function can be represented by a single condensed wave function with good angular momentum $0^+$. The maximum value amounts to 0.82 at $\vc{\beta}=\vc{\beta_2}\equiv (6.5\ {\rm fm}, 1.5\ {\rm fm})$ for the case of Volkov No. 2 force. The same calculation gives 0.88 at $\vc{\beta}=\vc{\beta_3}\equiv (7.5\ {\rm fm}, 1.5\ {\rm fm})$ for Volkov No. 1 force. This should be compared with the fact that the maximum squared overlap between the $0_2^+$ wave function obtained by Uegaki {\it et al.} \cite{carbon} and the Brink's $0^+$ wave function with a single three-alpha configuration was at most 0.5. This result gives further support for the interpretation that the $0_2^+$ state has a gas-like structure where three-alpha particles are condensed into an indentical $S$-wave orbit. 

We should keep in mind that the wave function ${\widehat \Phi}^{J=0}_{3\alpha}(\vc{\beta})$ of Eq. (\ref{eq:11}) necessarily contains compact components like the shell model state even if the size parameter $\vc{\beta}$ is large. Therefore the overlap $\langle {\widehat \Phi}^{{\rm N},J=0}_{3\alpha}(\vc{\beta}) | \Psi_{\lambda= 2}^{J=0}  \rangle$ even for large $\vc{\beta}=\vc{\beta_2}$ or $\vc{\beta}=\vc{\beta_3}$ suffers from  the effect of the compact components contained in ${\widehat \Phi}^{{\rm N},J=0}_{3\alpha}(\vc{\beta})$ with $\vc{\beta}=\vc{\beta_2}$ or $\vc{\beta_3}$. 
In order to calculate the overlap of $\Psi^{J=0}_{\lambda =2}$ with the gas-like state which does not contain the compact components, we need to subtract the compact components from ${\widehat \Phi}^{{\rm N},J=0}_{3\alpha}(\vc{\beta})$. 
The wave function $P^{J=0}_\bot(\vc{\beta_0}){\widehat \Phi}^{J=0}_{3\alpha}(\vc{\beta})$ is just such a kind of wave function in which compact components represented by ${\widehat \Phi}_{3\alpha}^{J=0}(\vc{\beta_0})$ are subtracted. As we mentioned above, the squared overlap of $\Psi^{J=0}_{\lambda =2}$ with $P^{J=0}_\bot(\vc{\beta_0}){\widehat \Phi}^{J=0}_{3\alpha}(\vc{\beta})$ can become as large as 0.97.

\begin{table*}
\begin{ruledtabular}
\begin{center}
\caption{Expectation values of kinetic energy, nucleon-nucleon interaction energy, and Coulomb interaction energy of the relative motions between $\alpha$ particles, denoted as $T-T_G$, $V_N$, and $V_C$, respectively. Internal contributions of the $\alpha$ particle are subtracted. Calculation is made for the $0^+_1$, $2^+_1$, $4_1^+$, and $0^+_2$ states for two cases of the effective two-nucleon force. Note that for the $4_1^+$ state, the spurious components of continuum states are subtracted by the method discussed in section \ref{subsec:analyses}. Units are in MeV.}
\begin{tabular}{ c c c c c c c c c} 
\label{tab:exp}
% \hline
 & \multicolumn{4}{c} {{\rm Volkov No.1} $M=0.575,\ E^{\rm th}_{3\alpha}=-81.01$\ MeV}
 & \multicolumn{4}{c} {{\rm Volkov No.2} $M=0.59 ,\ E^{\rm th}_{3\alpha}=-82.04$\ MeV} \\
% & \multicolumn{4}{}{$E^{\rm th}_{3\alpha}=-81.01$} 
% & \multicolumn{4}{}{$E^{\rm th}_{3\alpha}=-82.04$} \\
 \hline
$J^\pi$ &$\langle T-T_G \rangle$ &$\langle V_N \rangle$ & $\langle V_C \rangle$ & $E-E_{3\alpha}^{\rm th}$ 
 & $\langle T-T_G \rangle$ & $\langle V_N \rangle$ &$\langle V_C \rangle$ & $E-E_{3\alpha}^{\rm th}$ \\
 \hline
 $0^+_1$ & $56.4$ & $-68.8$ & $5.54$ & $-6.81$ & $51.2$ & $-64.2$ & $5.48$ & $-7.47$ \\
 $2^+_1$ & $60.7$ & $-70.6$ & $5.62$ & $-4.32$ & $55.6$ & $-65.8$ & $5.55$ & $-4.65$ \\
 $4_1^+$ & $71.9$ & $-76.6$ & $5.82$ & $1.14$  & $68.2$ & $-72.2$ & $5.77$ & 1.74 \\
 $0^+_2$ & $13.8$ & $-15.9$ & $3.23$ & 1.04 & $17.6$ & $-20.9$ & $3.58$ & 0.26
% \hline
\end{tabular}
\end{center}
\end{ruledtabular}
\end{table*}

We now discuss the ground rotational band states with $J^\pi=0_1^+$, $2_1^+$, and $4_1^+$. As we mentioned already, the binding energies of the ground rotational band states by our GCM are in good agreement with those by Uegaki {\it et al.} and Kamimura {\it et al.}, and therefore our GCM wave functions of these states are almost identical to their wave functions. It is not unnatural that our GCM wave functions could well represent these states, if we consider the fact that the normalized wave function of Eq. (\ref{eq:10}) goes to a shell model Slater determinant when $\beta_x=\beta_y=\beta_z \rightarrow 0$. However these states have a quite different structure from that of the $0_2^+$ state and cannot be considered to form a dilute $3\alpha$ cluster structure.

 In TABLE. \ref{tab:exp}, we give the kinetic energy, nucleon-nucleon interaction energy, and Coulomb interaction energy for each of the $0_1^+$, $2_1^+$, $4_1^+$, and $0_2^+$ states. We should note that the internal contributions of the alpha particles are subtracted. Kamimura {\it et al.} calculated the kinetic energy for each of the $0_1^+$, $2_1^+$, and $0_2^+$ states in Ref. \cite{carbon}, and our result well agrees with theirs. We see that for the $0_2^+$ state, the kinetic and nucleon-nucleon interaction energies are much smaller, compared with the case of the ground rotational band composed of the $0_1^+$, $2_1^+$, and $4_1^+$ states. The Coulomb interaction energy of the $0_2^+$ state is smaller than those of the ground rotational band states. These results are well understood by the fact that the $0_2^+$ state has a strongly developed $3\alpha$ cluster structure with low density where the inter-alpha binding is small. Simultaneously, it is indicated that the ground rotational band states have a compact structure with normal density, as is clearly shown by correspondingly small values of r.m.s radii of these states. 

\begin{figure}[htbp]
\begin{center}
\includegraphics[scale=.7]{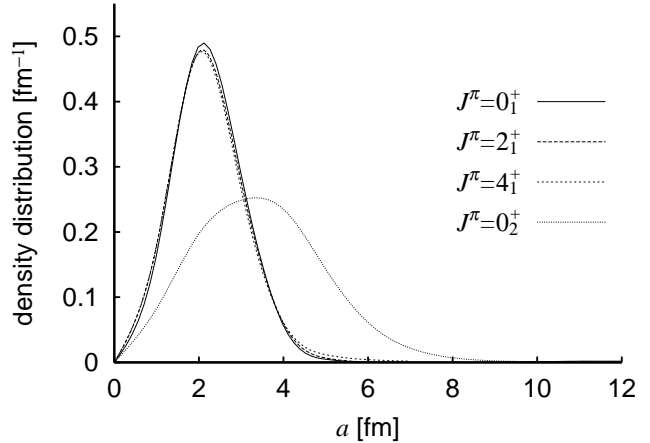}
\caption{Density distributions of the $0_1^+$, $2_1^+$, $4_1^+$, and $0_2^+$ states which are the expectation values of the density operator defined in Eq. (\ref{eq:dsty}). The adopted effective force is Volkov No. 2.}
\label{fig:1}
\end{center}
\end{figure}

 In FIG. \ref{fig:1}, we show the density distribution which is the expectation values of density operater defined as,
\begin{equation}
{\widehat \rho}(a) = \frac{1}{A} \sum_{i=1}^A \delta (|\vc{r}_i-\vc{X}_G|-a) .\label{eq:dsty}
\end{equation}
A brief explanation of how to derive the expectation values of the above density operator will be given in the APPENDIX. Obviously the wave function of the $0_2^+$ state leads to a low density structure with a long tail, while the states of the ground rotational band have compact structures with all almost the same density distribution. This result provides a strong support to the idea that the $0_2^+$ state can be described as a dilute density state forming a gas-like structure composed of three-alpha particles, as characterized by the wave function, Eq. (\ref{eq:10}). On the contrary, the ground rotational band states with $J^\pi=0_1^+$, $2_1^+$, and $4_1^+$ have a quantitatively different structure from the one of the $0_2^+$ state. Hence even though these states could be well represented by our GCM wave functions, they could never be described in terms of a Bose-condensation composed of three-alpha particles. 

 The ground band spectrum obtained by us as well as those by Uegaki {\it et al.}\cite{uegaki} and by Kamimura {\it et al.}\cite{kamimura} are quite compressed in comparison with experiment. This is a well-known problem of microscopic cluster model calculations. Takigawa and Arima \cite{takigawa} showed that the effect of the spin-orbit force taken into account by broken symmetry states with $[4431]$, $[4422]$, and so on introduced into the model space in addition to the cluster model states with maximum spatial symmetry $[444]$ can largely remedy this shortcoming of the cluster model. The pairing correlations may also have to be considered.

\subsection{ACCC study of $2_2^+$, $4_2^+$ and $0_2^+$}\label{subsec:resaccc}

\begin{figure}[htbp]
\begin{center}
\subfigure[]{
\label{fig:3(a)}
\includegraphics[scale=.85]{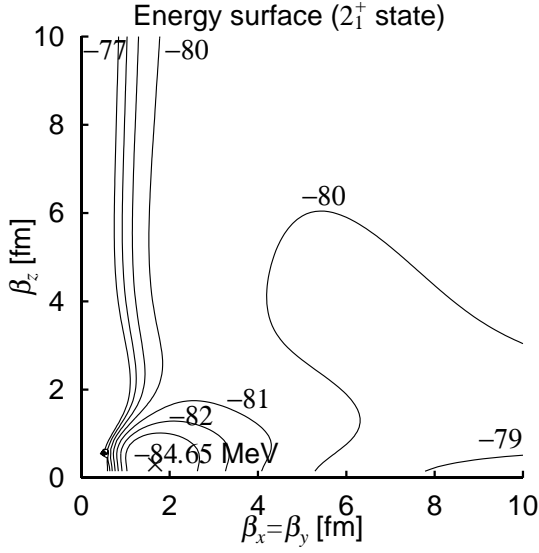}}
\subfigure[]{
\label{fig:3(b)}
\includegraphics[scale=.85]{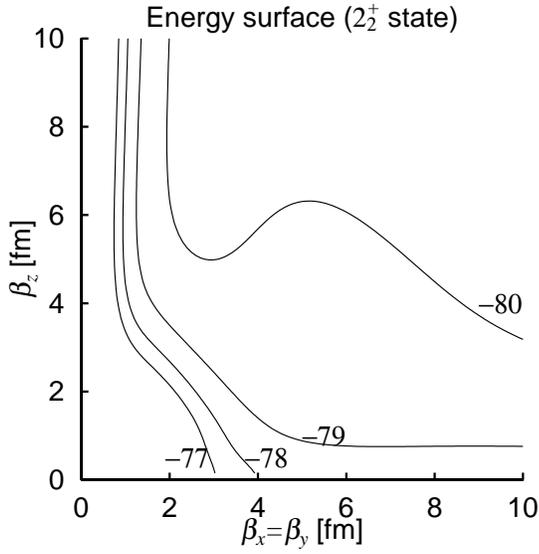}}
\caption{(a): Contour map of the energy surface of the $2^+$ state 
in the two-parameter space, $\beta_x(=\beta_y)$ and $\beta_z$. 
(b): Contour map of the energy surface corresponding to the $2^+$ 
state orthogonalized to the state at the minimum energy point in 
Fig. \ref{fig:3(a)}.
In both figures, the adopted effective nucleon force is Volkov No. 2, and numbers attached to the contour lines are binding energy values given in units of MeV.}
\end{center}
\end{figure}

The $2_2^+$ state is located at $2.6\pm 0.3$ MeV above the $3\alpha$ threshold with the $\alpha$-decay width $1.0\pm 0.3$ MeV. Therefore the bound state approximation used for the $0_2^+$ state, which is only $0.38$ MeV above the $3\alpha$ threshold, is no more useful for the $2_2^+$ state. This situation is shown in FIGs. \ref{fig:3(a)}, \ref{fig:3(b)}.

In FIG. \ref{fig:3(a)}, we give the contour map of the energy surface for the $2^+$ states in the parameter space, $\vc{\beta}$. We obtained the minimum energy point at $\vc{\beta}=\vc{\beta_0^D}\equiv(1.5\ {\rm fm}, 0.35\ {\rm fm})$. The optimum wave function is written as ${\widehat \Phi}_{3\alpha}^{J=2}(\vc{\beta_0^D})$, which corresponds to the $2_1^+$ state, since the binding energy, $-84.65$ MeV is similar to the correct binding energy of the $2_1^+$ state, $-86.70$ MeV, as indicated in TABLE \ref{tab:1}. In order to look for the $2_2^+$ state, we define a projection operator,
\begin{equation}
P^{J=2}_\bot(\vc{\beta_0^D})=1-\big|{\widehat \Phi}_{3\alpha}^{N,J=2}(\vc{\beta_0^D}) \big\rangle \big\langle {\widehat \Phi}_{3\alpha}^{N,J=2}(\vc{\beta_0^D})\big|. \label{eq:26}
\end{equation}
Using this, we calculated the $2^+$ energy surface of the orthogonalized states, $P^{J=2}_\bot(\vc{\beta^D_0}){\widehat \Phi}_{3\alpha}^{J=2}(\vc{\beta})$, to the minimum energy state. The contour map is shown in FIG. \ref{fig:3(b)}. There appears no minimum point in this map. This is because the $2_2^+$ state lies close to the top of the Coulomb and centrifugal barriers. This situation of the $2_2^+$ state is in marked contrast to that of the $0_2^+$ state for which, as shown in Ref. \cite{cbec}, we could see a clear minimum point in the contour map of the energy surface for the $0^+$ states calculated in this way. This is because the $0_2^+$ state lies well below the top of the Coulomb barrier. Therefore, the $2_2^+$ state has to be correctly treated as a resonance state.

\begin{figure}[htbp]
\begin{center}
\subfigure[]{
\label{fig:4(a)}
\includegraphics[scale=.7]{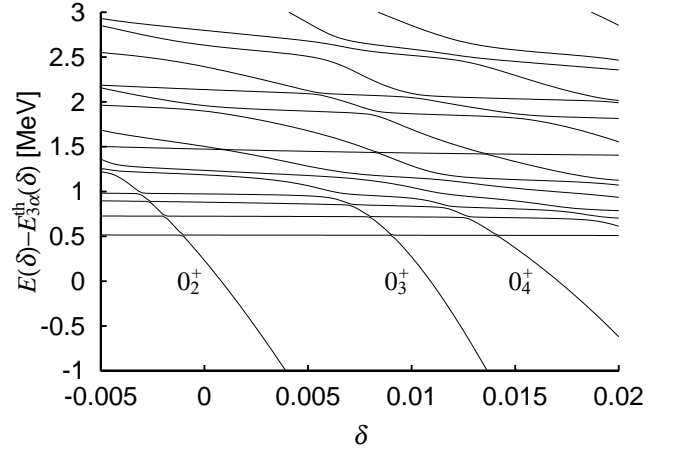}}
\subfigure[]{
\label{fig:4(b)}
\includegraphics[scale=.7]{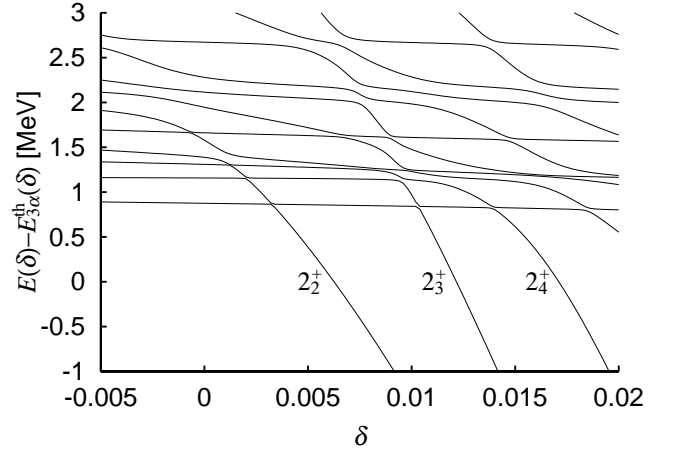}}
\subfigure[]{
\label{fig:4(c)}
\includegraphics[scale=.7]{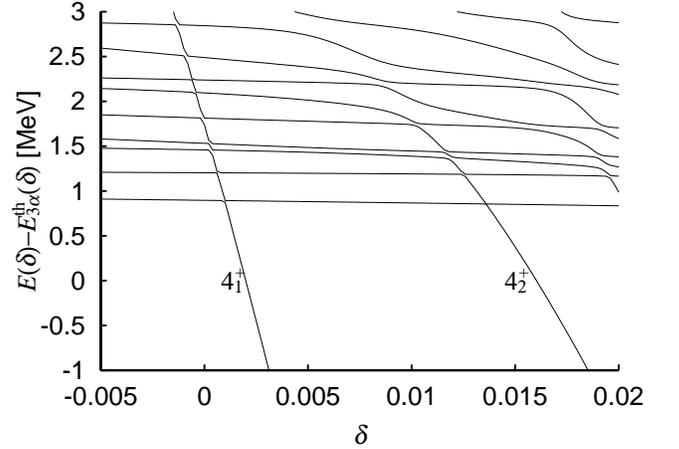}}
\caption{The energy eigenvalues of the Hill-Wheeler equation, Eq. (\ref{eq:27}), plotted as a function of the coupling constant, $\delta$. (a): For $J^\pi =0^+$.  (b): For $J^\pi =2^+$. (c): For $J^\pi =4^+$. The adopted effective nucleon force is Volkov No. 2 for all cases.}
\end{center}
\end{figure}

In FIGs.\ref{fig:4(a)}-\ref{fig:4(c)}, we plot the energy eigenvalues of the ACCC Hamiltonian $H^\prime$ of Eq. (\ref{eq:14}) obtained by solving the Hill-Wheeler equation, Eq. (\ref{eq:27}), for $J^\pi=0^+, 2^+$, and $4^+$.
The solutions of the above equation of motion depend on the coupling constant, $\delta$, and the energy eigenvalues are plotted as a function of $\delta$. It is seen that resonance solutions dive under the line of the $3\alpha$ threshold, $E=0$, as $\delta$ increases. On the other hand, the continuum solutions, which are insensitive to variations of $\delta$ and look just like horizontal straight lines on the figures, stay above the line of the $3\alpha$ threshold. This is how resonance states can be separated from continuum states. It is to be noted that around $\delta=0$, an undesirable mixing of resonance and continuum states occurs, since the Hill-Wheeler equation, Eq. (\ref{eq:27}), has not been solved with proper boundary conditions. Therefore the correct resonance states cannot be clearly identified. As characteristic features of these figures, the trajectory of the $0_2^+$ state as well as the $4_1^+$ state can be clearly traced, even when the coupling constant, $\delta$, decreases to zero, while the trajectories of the other resonance states, for instance, that of the $0_3^+$ state get obscure when $\delta$ becomes less than about $0.01$. As a result, the $0_3^+$ state cannot be distinguished from a few neighboring states at $\delta = 0$. It follows that it is a very good approximation to solve the $0_2^+$ state within the framework of the bound state approximation as practiced in the previous subsection, while for the $2_2^+$, $0_3^+$, and $4_2^+$ states with non negligible widths, it is necessary to go beyond the framework of the bound state approximation. In order to solve those resonance states, the method of ACCC is applied here. 

\begin{figure}[htbp]
\begin{center}
\subfigure[]{
\label{fig:5(b)}
\includegraphics[scale=.7]{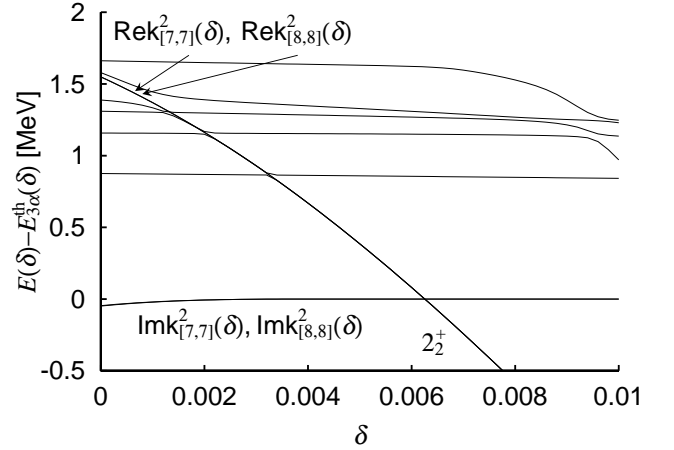}}
\subfigure[]{
\label{fig:5(c)}
\includegraphics[scale=.7]{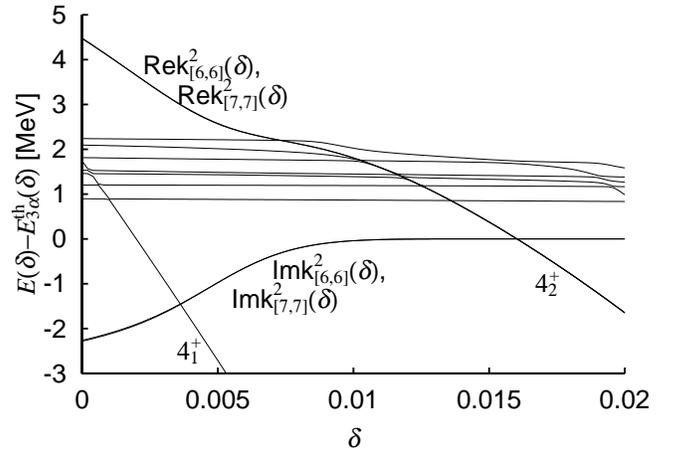}}
\caption{(a): Analytically continued complex energy functions for the $2_2^+$ state are drawn by superposing them to FIG. \ref{fig:4(b)}. Note that $k^2_{[7,7]}$ and $k^2_{[8,8]}$ are converged, so that their trajectories can be hardly distinguished from one another in this figure. (b): Analytically continued complex energy functions for the $4_2^+$ state are drawn by superposing them to FIG. \ref{fig:4(c)}. Note that $k^2_{[6,6]}$ and $k^2_{[7,7]}$ are converged, so that their trajectories can be hardly distinguished from one another in this figure.}
\end{center}
\end{figure}

In Figs.\ref{fig:5(b)}, \ref{fig:5(c)}, we show the converged complex energy functions for the $2_2^+$ and $4_2^+$ states. The energy functions of resonance states have the form of Eqs. (\ref{eq:16}) and (\ref{eq:17}). As already mentioned in section \ref{subsec:accc}, once the coefficients of the Pad\'e rational function are determined by the use of the calculated energy eigenvalues underneath the $3\alpha$ threshold, the position and total decay width can be derived by extrapolating the complex energy functions of the coupling constant, $\delta$, to $\delta=0$. For the $2_2^+$, and $4_2^+$ states, we made use of energy region down to about 3 MeV starting from several tens of keV below the $3\alpha$ threshold in order to determine the coefficients of the Pad\'e approximant. 

 In FIGs. \ref{fig:5(b)}, \ref{fig:5(c)}, the calculated stable trajectories for a few kinds of Pad\'e approximants labeled as $[N, M]$ ($N=M$) defined in Eq. (\ref{eq:16}) are drawn by superposing them on FIGs. \ref{fig:4(b)}, \ref{fig:4(c)}, respectively. We see that the real parts of all calculated trajectories retrace, even above the threshold, the trajectories in FIGs. \ref{fig:4(b)}, \ref{fig:4(c)}. It seems that we need to adopt the Pad\'e approximant of $N=M \geq 6$ for these states in order to correctly retrace the real parts of the complex energy functions, especially above the threshold, the energy eigenvalues obtained by the bound state approximation. On the other hand, we applied the Pad\'e approximations of $N=M \geq 9$ to the $2_2^+$ state, and $N=M \geq 8$ to the $4_2^+$ state, but we did not adopt the results of these applications because they appear to be ill-behaved or divergent. It should be noted that the two kinds of analytically continued Pad\'e approximants of different $N=M$ values for the each state on FIGs. \ref{fig:5(b)}, \ref{fig:5(c)} almost coincide so that we can hardly distinguish them from one another.

 For the $2_2^+$ state, FIG. \ref{fig:5(b)} shows the energy position, ${\rm Re}(E(\delta=0)-E_{3\alpha}^{\rm th}(\delta=0))=1.55$ MeV, and total decay width, $\Gamma=-2\cdot {\rm Im}(E(\delta=0)-E_{3\alpha}^{\rm th}(\delta=0))=0.09$ MeV, when the Volkov No. 2 force is adopted. The calculated binding energy, $1.55$ MeV, is quite a bit smaller than the observed value, $2.6\pm 0.3$ MeV. According to the lower binding energy which makes the decay penetrability smaller, the total decay width is also much smaller, i.e. 0.09 MeV, than the observed value, $1.0\pm 0.3$ MeV. When we adopt the Volkov No. 1 force, the energy position is calculated as, ${\rm Re}(E(\delta=0)-E_{3\alpha}^{\rm th}(\delta=0))=2.1$ MeV. According to the higher energy position, the calculated total decay width is larger than the one of the Volkov No. 2 case and is, $\Gamma=-2\cdot {\rm Im}(E(\delta=0)-E_{3\alpha}^{\rm th}(\delta=0))=0.64$ MeV, which quite well reproduces the experimental data. The sensitive dependence of the penetrability on the energy position can be seen by adjusting the value of $\delta$ and by making the whole system a little less bound, so that the real part of energy function takes a reasonable value. When $\delta=-0.0035$, the real part of energy function, Re($E_{[8,8]}(\delta=-0.0035)-E_{3\alpha}^{\rm th}(\delta=-0.0035))=2.0$ MeV, and $\Gamma=0.61$ MeV. At this value of $\delta$, the threshold energy of $^8{\rm Be}+\alpha$ turns to be $0.172$ keV measured from the three alpha threshold energy, though at $\delta=0$ the $^8$Be$+\alpha$ threshold energy is $-0.244$ MeV. When we adopt the Volkov No. 1 force, the threshold energy of $^8{\rm Be}+\alpha$ at $\delta=0$ is 0.191 MeV measured from the three alpha breakup threshold energy. Consequently, we can say that for both effective nucleon forces, the resonance energy and width of the $2_2^+$ state are mutually consistent and in good agreement with the recently observed experimental data. 

There is another way to adjust the present unrealistic penetrability in the case of the Volkov No. 2 force to a more reasonable value. It is using the $R$-matrix theory in which the decay width is expressed as $\Gamma=2P_l\cdot \gamma^2$, as given in Eq. (\ref{eq:28}).
% a=6 theta=0.39 gamma'=gamma*P'/P=0.976 for v1 22+
% a=6 theta=0.062  gamma'=gamma*P'/P=0.15
From this relation, we can estimate the total decay width corresponding to the observed energy position of the $2_2^+$ state, $2.5 \pm 0.3$ MeV above the $^8{\rm Be}+\alpha$ breakup threshold energy which is 0.092 MeV above the three alpha breakup threshold. The estimation is made through the following relation:

\begin{equation}
\Gamma^\prime =\frac{P_l(\bar{\epsilon})}{P_l(\epsilon(\delta=0))}\Gamma(\delta=0). \label{eq:29}
\end{equation}

The calculated total dacay width, $\Gamma^\prime$ is 0.24 MeV when the channel radius $a=6$ fm, which is still smaller than the observed value. It is to be noted that the above expression of penetrability is an approximate one and only applicable to the case of a narrow resonance width. Since the observed width of the $2_2^+$ state which is $1.0\pm 0.3$ MeV, is no longer narrow, the use of this expression of penetrability may lead to an unsatisfactory result.

 In the present work, we do not argue about the $2_3^+$ and $2_4^+$ states which are shown in Fig. \ref{fig:4(b)}. A recent experiment assigns a $2^+$ state at the excitation energy, $\sim 14$ MeV with the width, $\sim 1$ MeV in the preliminary analysis \cite{fynbo}. Theoretical predictions that a $2^+$ state exsists in this energy region with the width, $\sim 1$ MeV are also reported \cite{kurokawa, csoto}.

 In section \ref{subsec:accc}, we devised an approximate method to calculate the width which is much smaller than the energy measured from the threshold within the framework of ACCC method. We calculated the width of the $0_2^+$ state by using Eq. (\ref{eq:penetrability}), $\Gamma(\delta=0)$$\approx$$ (P_{l=0}(\bar{\epsilon}) / P_{l=0}(\epsilon(\delta_z)) )\Gamma(\delta_z)$, where the observed decay energy $\bar{\epsilon}$$=$$0.288$ MeV. When $\delta_z$$=$$-0.0035$, the calculated width $\Gamma(\delta=0)$$\approx$$7.0\times 10^{-6}$ MeV, where the decay energy and width are $\epsilon(\delta_z)$$=$$0.643$ MeV and $\Gamma(\delta_z)$$=$$0.012$ MeV, respectively. When $\delta_z$$=$$-0.005$, the calculated width $\Gamma(\delta=0)$$\approx$$9.9\times 10^{-6}$ MeV, where the decay energy and width are $\epsilon(\delta_z)$$=$$0.790$ MeV and $\Gamma(\delta_z)$$=$$0.064$ MeV, respectively. The adopted channel radius is $a$$=$$6$ fm in this calculation. The obtained values are consistent with the observed width $8.7\pm 2.7$ eV \cite{ajze}.

 In FIG. \ref{fig:5(c)}, the energy function of the $4_2^+$ state using the Volkov No. 2 force is drawn. The obtained binding energy mesured from the threshold and the total decay width are 4.5 MeV and 4.6 MeV, respectively. When we adopt the Volkov No. 1 force, we obtain 5.4 MeV and 4.0 MeV for the energy and the width, respectively. In the case of the Volkov No. 1 force, Uegaki {\it et al.} predicted in the Ref. \cite{uegaki} the exsistence of the $4_2^+$ state that is, however, experimentaly unknown. In spite of the fact that we obtained a unique and stable trajectory for the $4_2^+$ state for both of the adopted effective nuclear forces, we, however, could not make a decisive argument about whether the obtained $4_2^+$ state possesses a correct physical meaning. The reason for this will be clarified in the following subsection.

\subsection{Analysis of wave functions of the $2_2^+$- and $4_2^+$- states}\label{subsec:analyses}

 We have seen that our wave function of the $2_2^+$ state reasonably reproduces the observed binding energy and alpha decay width, with the help of the ACCC method. In this subsection, we analyse the wave function of the $2_2^+$ state. Since the method of ACCC, unfortunately, does not provide the wave function of the resonance state, we utilize the GCM wave functions for the ACCC Hamiltonian $H^\prime(\delta)$ in order to estimate the correct $2_2^+$ wave function. 

 FIG. \ref{fig:4(b)} shows that the binding energy of the $2_2^+$ state coincides with the threshold energy when $\delta=0.00626$. Let $\Psi ^J_\lambda (\delta)$ be a solution, Eq. (\ref{eq:30}), of the Hill-Wheeler equation, Eq. (\ref{eq:27}). For $\delta \leq 0.0062$, as long as we can safely trace the clearly visible trajectory of the resonance state corresponding to the $2_2^+$ state shown in FIG. \ref{fig:4(b)}, we will be able to regard the wave function $\Psi^{J=2}_\lambda (\delta)$ on the trajectory of the $2_2^+$ state as consisting of an almost pure resonance component for the ACCC Hamiltonian $H^\prime(\delta)$. Below we give arguments for this conjecture: as already mentioned, the horizontal straight lines in Fig. \ref{fig:4(b)} express the trajectories of the continuum states. We can check this fact by calculating the density distribution $\rho(a)$ of any state on the continuum-state trajectories, which shows that $\rho(a)\lesssim 0.05$ for $a\leq 10$ fm. The coupling of the resonance state with the continuum state is the strongest at the point where the resonance-state trajectory crosses the continuum-state trajectory. The coupling of the resonance state with the continuum state becomes weaker as we go away from the crossing point along the resonance-state trajectory. When we arrive at a point on the resonance-state trajectory where the effect of the level-level repulsion between the resonance-state and continuum-state trajectories is no more existent, we may consider that the mixing of the continuum state into the resonance state is very small. 

\begin{figure}[htbp]
\begin{center}
\includegraphics[scale=.7]{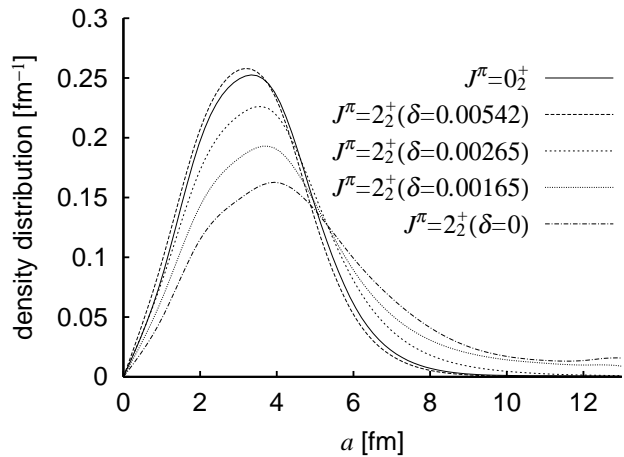}
\caption{Density distributions of the wave functions, $\Psi ^{J=2}_{\lambda =2} ($$\delta$$=$$0.00542)$, $\Psi ^{J=2}_{\lambda =3} ($$\delta$$=$$0.00265)$, $\Psi ^{J=2}_{\lambda =4} ($$\delta$$=$$0.00165)$, and $\Psi ^{J=2}_{\lambda =6} ($$\delta$$=$$0)$, together with that of the $0_2^+$ state given in the section, \ref{subsec:gcm}. The density operator is defined in Eq. (\ref{eq:dsty}). Volkov No. 2 force is adopted.}
\label{fig:dy2ana}
\end{center}
\end{figure}

In FIG. \ref{fig:dy2ana} we show the density distributions $\rho(a)$ of four states on the $2_2^+$ resonance trajectory, $\psi_{\rm I}\equiv \Psi^{J=2}_{\lambda=2}(\delta$$=$$0.00542)$, $\psi_{\rm II}\equiv \Psi^{J=2}_{\lambda=3}(\delta$$=$$0.00265)$, $\psi_{\rm III}\equiv \Psi^{J=2}_{\lambda=4}(\delta$$=$$0.00165)$, and $\psi_{\rm IV}\equiv \Psi^{J=2}_{\lambda=6}(\delta=0)$. As is seen in FIG. \ref{fig:4(b)}, the first two states, $\psi_{\rm I}$ and $\psi_{\rm II}$, are out of influence of the crossing with continuum trajectories. However, the last state, $\psi_{\rm IV}$, seems to contain non-negligible amount of the continuum state component, because this state is located near the crossing point between the $2_2^+$ trajectory and the fourth lowest continuum trajectory with large mutual repulsion. In addition, it may be said that the state, $\psi_{\rm III}$, has little mixing with the second and third continuum states, since the state is closely located between the second and third crossing points. For the sake of comparison, we also show in FIG. \ref{fig:dy2ana} the density distribution of the $0_2^+$ state which we already showed in FIG. \ref{fig:1}. We note that the density distributions of the first two states, $\psi_{\rm I}$ and $\psi_{\rm II}$, are confined inside a finite spatial region. Actually the integration of $\rho(a)$ 
\begin{equation}
I(R)=\int_0^R \rho(a) da
\end{equation}
which should be unity for $R=\infty$, $I(R=\infty)=1$, becomes almost unity already for finite $R$ except for the last two states, $\psi_{\rm III}$ and $\psi_{\rm IV}$. Such finite values of $R$ are $8$ fm for the $0_2^+$ state and $\psi_{\rm I}$, and $11$ fm for $\psi_{\rm II}$. In the cases of $\psi_{\rm III}$ and $\psi_{\rm IV}$, $I(R$$=$$13$ fm)$=$$0.96$ and $0.90$, respectively, and $\rho(a)$ is not damped away even beyond $a=15$ fm.

\begin{figure}[htbp]
\begin{center}
\subfigure[]{
\label{fig:ana1}
\includegraphics[scale=.9]{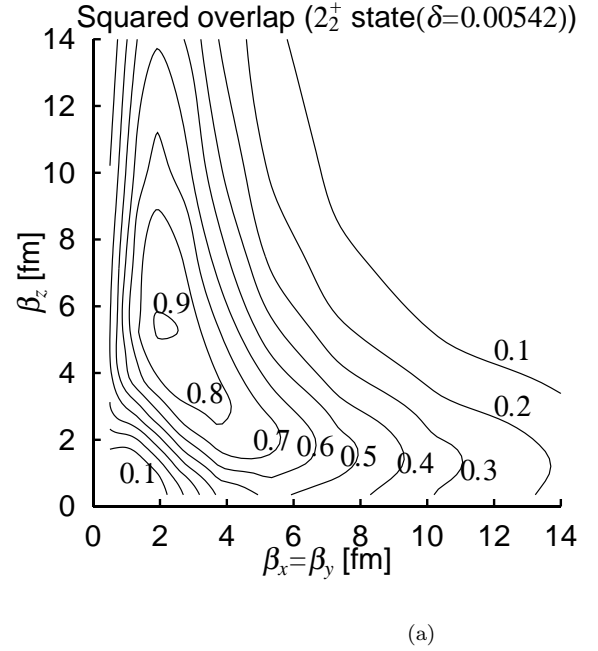}}
\subfigure[]{
\label{fig:ana2}
\includegraphics[scale=.9]{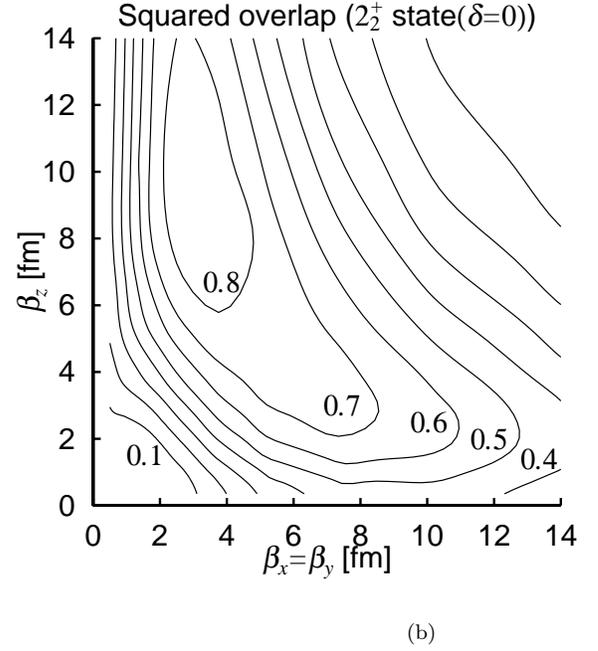}}
\caption{(a): Contour map of the squared overlap, $| \langle {\widehat \Phi}_{3\alpha}^{{\rm N},J=2}(\vc{\beta}) | \Psi _{\lambda=2}^{J=2}($$\delta$$=$$0.00365) \rangle |^2$, in the two-parameter space, $\beta_x(=\beta_y)$ and $\beta_z$. (b): Contour map of the squared overlap, $| \langle {\widehat \Phi}_{3\alpha}^{{\rm N},J=2}(\vc{\beta}) | \Psi _{\lambda=6}^{J=2}($$\delta$$=$$0) \rangle |^2$, in the two-parameter space, $\beta_x(=\beta_y)$ and $\beta_z$. Volkov No. 2 force is adopted.}
\end{center}
\end{figure}

The reason for the choice of $\delta=0.00542$ of the first state $\psi_{\rm I}\equiv \Psi^{J=2}_{\lambda=2}(\delta=0.00542)$ is to make its excitation energy measured from the $3\alpha$ threshold, $E(\delta)-E^{\rm th}_{3\alpha}(\delta)$, be equal to that of the $0_2^+$ state which is $0.26$ MeV for our present choice of Volkov No. 2 force. We see that the density distribution of this state, $\psi_{\rm I}$, is almost the same as that of the $0_2^+$ state. This result implies that if we pull down the excitation energy of the $2_2^+$ state to the same value as that of the $0_2^+$ state, the density distribution of the $2_2^+$ state becomes almost identical to that of the $0_2^+$ state. In FIG. \ref{fig:ana1}, we show the squared overlap between $\psi_{\rm I}$ and the single $2^+$ condensate wave function ${\widehat \Phi}^{{\rm N}, J=2}_{3\alpha}(\vc{\beta})$, i.e. $|\langle {\widehat \Phi}^{{\rm N}, J=2}_{3\alpha}(\vc{\beta})| \psi_{\rm I}\rangle|^2$. We see that the maximum value of the squared overlap is more than $90$ \%. Here we should recall that in the case of the $0_2^+$ state the maximum value of such squared overlap with single $0^+$ condensate wave function ${\widehat \Phi}^{{\rm N}, J=0}_{3\alpha}(\vc{\beta})$ amounts to $82$ \% for the Volkov No. 2 force and to $88$ \% for the Volkov No. 1 force. We further calculated the squared overlap of $\psi_{\rm I}$ with $P_{\bot}^{J=2}(\vc{\beta}_0^{D}){\widehat \Phi}_{3\alpha}^{J=2}$ which is the projection operator defined in Eq. (\ref{eq:26}). We found the maximum value to be more than 95 \%. In the case of the $0_2^+$ state, a similar calculation of the squared overlap with $P^{J=0}_\bot(\vc{\beta}_0)\widehat{\Phi}_{n\alpha}^{J=0}(\vc{\beta}_1)$ gives a value as large as 97 \%, where $\vc{\beta}_1$$=$$(5.7\ {\rm fm}, 1.5\ {\rm fm})$, as mentioned in section \ref{subsec:gcm}. We thus can say that the wave function $\psi_{\rm I}$ has a very similar structure as the $0_2^+$ state of Bose-condensate character.

We see in FIG. \ref{fig:dy2ana} that the density distribution $\rho(a)$ of the second state, $\psi_{\rm II}$$\equiv$$\Psi^{J=2}_{\lambda=3}(\delta$$=$$0.00265)$, is slightly pushed outwards compared with that of $\psi_{\rm I}$, which we consider to be natural because the excitation energy of $\psi_{\rm II}$ measuredd from the $3\alpha$ threshold, $E(\delta)-E^{\rm th}_{3\alpha}(\delta)$, is $1.01$ MeV and hence is higher than that of $\psi_{\rm I}$ which is $0.26$ MeV. We also calculated the squared overlap between $\psi_{\rm II}$ and the single $2^+$ condensate wave function ${\widehat \Phi}^{{\rm N}, J=2}_{3\alpha}(\vc{\beta})$, $|\langle {\widehat \Phi}^{{\rm N}, J=2}_{3\alpha}(\vc{\beta})| \psi_{\rm II}\rangle|^2$, and we found that the maximum value of the squared overlap is approximately $90$ \%.

In the cases of the third and fourth states, $\psi_{\rm III}\equiv \Psi^{J=2}_{\lambda=4}(\delta_{\rm III})$ with $\delta_{\rm III}$$=$$0.00165$, and $\psi_{\rm IV}\equiv \Psi^{J=2}_{\lambda=6}($$\delta$$=$$0)$, as we mentioned already, there exist mixings of continuum states into these states. FIG. \ref{fig:4(b)} tells us that the continuum states which mix into $\psi_{\rm IV}$ seem to be dominantly the fourth lowest continuum state which we denote as $\varphi_4(\delta$$=$$0)$. But there also may be small mixing of the third and fifth continuum states, $\varphi_3($$\delta$$=$$0)$ and $\varphi_5($$\delta$$=$$0)$, into $\psi_{\rm IV}$. In the case of the third state, $\psi_{\rm III}$, there may be small mixing of the second and third continuum states, $\varphi_2(\delta_{\rm III})$ and $\varphi_3(\delta_{\rm III})$. Thus $\psi_{\rm III}$ and $\psi_{\rm IV}$ can be expressed as

\begin{eqnarray}
\psi_{\rm III} &=& C_1^{\rm III} \phi_{\rm R}(\delta_{\rm III}) + C_2^{\rm III} \varphi_2(\delta_{\rm III}) + C_3^{\rm III} \varphi_3(\delta_{\rm III}) , \nonumber \\
\psi_{\rm IV} &=& C_1^{\rm IV} \phi_{\rm R}(\delta=0) + C_2^{\rm IV} \varphi_4(\delta=0) \nonumber \\
 &&\hspace{0.8cm} + C_3^{\rm IV} \varphi_3(\delta=0) + C_4^{\rm IV} \varphi_5(\delta=0) .
\end{eqnarray}
Here, $\phi_{\rm R}(\delta_{\rm III})$ and $\phi_{\rm R}($$\delta$$=$$0)$ are just the resonance states which we want to extract from $\psi_{\rm III}$ and $\psi_{\rm IV}$, respectively.

\begin{figure}[htbp]
\begin{center}
\includegraphics[scale=.7]{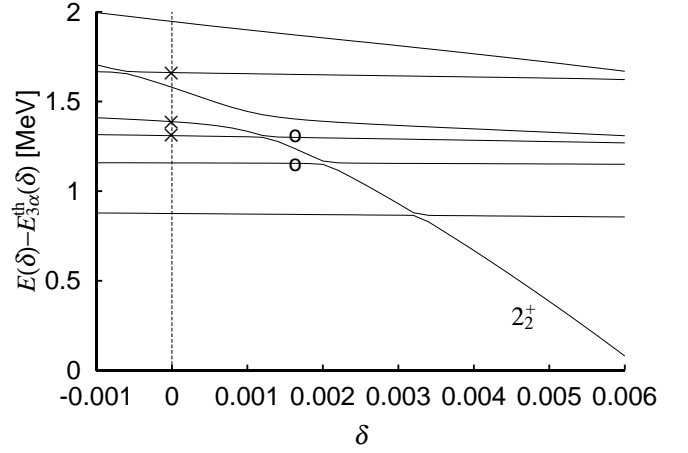}
\caption{The energy eigenvalues for $J^\pi =2^+$ of the Hill-Wheeler equation, Eq. (\ref{eq:27}), ploted as a function of the coupling constant, $\delta$. $\Psi^{J=2}_{\lambda=4}(\delta$$=$$0)$, $\Psi^{J=2}_{\lambda=5}(\delta$$=$$0)$, and $\Psi^{J=2}_{\lambda=7}(\delta$$=$$0)$ are marked by crosses in order from the bottom, and $\Psi^{J=2}_{\lambda=3}(\delta_{\rm III}$) and $\Psi^{J=2}_{\lambda=5}(\delta_{\rm III}$) by open circles. Volkov No. 2 force is adopted.}
\label{fig:ext}
\end{center}
\end{figure}

The extraction of these resonance states, $\phi_{\rm R}(\delta_{\rm III})$ and $\phi_{\rm R}($$\delta$$=$$0)$, can be made in the following way. We explain the procedure in the case of $\phi_{\rm R}(\delta$$=$$0)$. First we note that the three eigen-energy states of GCM, $\Psi^{J=2}_{\lambda=4}(\delta$$=$$0)$, $\Psi^{J=2}_{\lambda=5}(\delta$$=$$0)$, and $\Psi^{J=2}_{\lambda=7}(\delta$$=$$0)$, which are the $\delta$$=$$0$ states on the third, fourth, and fifth lowest continuum trajectories, respectively, and which are marked by crosses on the $\delta$$=$$0$ vertical dashed line in FIG. \ref{fig:ext}, can also be expressed by a linear combination of four states, $\phi_{\rm R}(\delta$$=$$0)$, $\varphi_4(\delta$$=$$0)$, $\varphi_3(\delta$$=$$0)$, and $\varphi_5(\delta$$=$$0)$, just like $\psi_{\rm IV}$$=$$\Psi^{J=2}_{\lambda=6}(\delta$$=$$0)$. This means that $\phi_{\rm R}(\delta$$=$$0)$ can be expressed by a linear combination of four GCM eigen-states
\begin{equation}
\phi_{\rm R}(\delta=0) = \sum_{\lambda=4}^7 D^{\rm IV}_\lambda \Psi^{J=2}_\lambda (\delta=0).
\end{equation}
Since the rms radius of $\phi_{\rm R}(\delta$$=$$0)$ should be much smaller than the rms radius of any of the three continuum states, $\varphi_4(\delta$$=$$0)$, $\varphi_3(\delta$$=$$0)$, and $\varphi_5(\delta$$=$$0)$, the coefficients $\{ D^{\rm IV}_\lambda \}$ of the above equation should satisfy that the expression 
\begin{eqnarray}
&&\Big\langle \sum_{\lambda=4}^7 D^{\rm IV}_\lambda \Psi_\lambda^{J=2}(\delta=0) \Big| \sum_{j=1}^{12} r_j^2 \Big| \sum_{\lambda=4}^7 D^{\rm IV}_\lambda \Psi_\lambda^{J=2}(\delta=0) \Big\rangle  \nonumber \\ 
&& 
\end{eqnarray}
takes the smallest value possible.
This condition determines the coefficients $\{D^{\rm IV}_\lambda \}$ and hence $\phi_{\rm R}(\delta$$=$$0)$. Similarly, we can extract the resonance state $\phi_{\rm R}(\delta_{\rm III})$ by linearly combining the GCM eigen-states, $\Psi^{J=2}_{\lambda=3}(\delta_{\rm III})$, $\psi_{\rm III}$$=$$\Psi^{J=2}_{\lambda=4}(\delta_{\rm III})$, and $\Psi^{J=2}_{\lambda=5}(\delta_{\rm III})$. The calculated squared overlap, $|\langle \phi_{\rm R}(\delta$$=$$0)|\psi_{\rm IV}\rangle|^2$$=$$|C^{\rm IV}_1|^2$$=$$|D_{\lambda=6}^{\rm IV}|^2$ and $|\langle \phi_{\rm R}(\delta_{\rm III})|\psi_{\rm III}\rangle|^2$$=$$|C^{\rm III}_1|^2$$=$$|D_{\lambda=4}^{\rm III}|^2$ is 0.811 and 0.967, respectively. This means that the degrees of mixing of continuum states into $\psi_{\rm IV}$ and $\psi_{\rm III}$ are around 19 \% and 3 \%, respectively. We have checked by using this technique that the two states, $\psi_{\rm II}$ and $\psi_{\rm I}$, are never influenced by their neighboring crossing points. 

In TABLE. \ref{tab:22+} we show various properties of the four resonance wave functions along the $2_2^+$ trajectory, $\psi_{\rm I}$, $\psi_{\rm II}$, $\phi_{\rm R}(\delta_{\rm III})$, and $\phi_{\rm R}(\delta$$=$$0)$. We have checked that the energies of these four resonance states (namely the expectation values of the Hamiltonian $H^\prime (\delta)$ by these four resonance states) measured from the $3\alpha$ threshold, $E(\delta)$$-$$E^{\rm th}_{3\alpha}(\delta)$, lie almost completely on the curve of the ACCC real energy for the $2_2^+$ state given in FIG. \ref{fig:5(b)}. This result shows the high reliability of our extracted resonance wave functions of $\phi_{\rm R}(\delta_{\rm III})$, and $\phi_{\rm R}(\delta$$=$$0)$. We see that the rms radius of the resonance state becomes larger as the excitation energy measured from the $3\alpha$ threshold becomes higher from $\psi_{\rm I}$ to $\phi_{\rm R}(\delta$$=$$0)$.

The resonance state $\phi_{\rm R}(\delta$$=$$0)$ is our approximate wave function of the $2_2^+$ state. Since the excitation energy of this state measured from the $3\alpha$ threshold is quite a bit higher than that of the $0_2^+$ state, its rms radius is much larger than that of the $0_2^+$ state unlike for the case of $\psi_{\rm I}$. However, this state preserves the same character as the $\psi_{\rm I}$ state that has a very similar structure to the $0_2^+$ state of Bose-condensate character. It is shown in FIG. \ref{fig:ana2} where there is displayed the squared overlap between $\phi_{\rm R}(\delta$$=$$0)$ and the single $2^+$ condensate wave function ${\widehat \Phi}^{{\rm N}, J=2}_{3\alpha}(\vc{\beta})$, namely $|\langle {\widehat \Phi}_{3\alpha}^{{\rm N}, J=2}(\vc{\beta})|\phi_{\rm R}(\delta$$=$$0\rangle|^2$. We see that the maximum value of the squared overlap is more than $88$ \%. We further calculated the squared overlap of $\phi_{\rm R}(\delta$$=$$0)$ with $P_{\bot}^{J=2}(\vc{\beta}_0^{D}){\widehat \Phi}_{3\alpha}^{J=2}$ which is the projection operator defined in Eq. (\ref{eq:26}). The maximum value is more than 90 \%.

\begin{table*}[htbp]
\begin{ruledtabular}
\begin{center}
\caption{Expectation values of kinetic energy, nucleon-nucleon interaction energy, and Coulomb interaction energy of the relative motions between $\alpha$ particles, denoted as $T-T_G$, $V_N$, and $V_C$, respectively. Internal contributions of the $\alpha$ particle are subtracted. Calculation is made for the $2^+_2$ state by using Volkov force No. 2 and various values of the coupling constant $\delta$. When $\delta$$=$0, 0.00165, 0.00265, 0.00542, the wave functions, $\phi_{\rm R}(\delta=0)$, $\phi_{\rm R}(\delta_{\rm III})$, $\Psi ^{J=2}_{\lambda =3} (\delta=0.00265)$, and $\Psi ^{J=2}_{\lambda =2} (\delta=0.00542)$ are adopted, respectively. Units are in MeV except for $R_{\rm r.m.s}$ whose unit is in fm.}\label{tab:22+}
\begin{tabular}{c c c c c c c c}
$J^\pi$&$\delta$ & $\langle T$$-$$T_G \rangle$ & $\langle V_N \rangle$ & $\langle V_C \rangle $ & $E(\delta)$$-$$E_{3\alpha}^{\rm th}(\delta)$ & $E_{3\alpha}^{\rm th}(\delta)$ &$R_{\rm r.m.s}$  \\
\hline
                         &0       & $9.84$ & $-11.1$ & $2.80$ & $1.54$ & $-82.04$ & $5.40$  \\ 
                         &0.00165 & $12.5$ & $-14.2$ & $3.00$ & $1.23$ & $-83.24$ & $5.39$  \\ 
\raisebox{1.5ex}[0cm][0cm]{$2_2^+$}  &0.00265 & $15.9$ & $-18.3$ & $3.35$ & $1.01$ & $-83.96$ & $4.32$  \\ 
                         &0.00542 & $21.3$ & $-24.8$ & $3.72$ & $0.26$ & $-85.96$ & $3.68$  \\ \hline
  $0_2^+$                &0       & $17.6$ & $-20.9$ & $3.58$ & $0.26$ & $-82.04$ & $3.83$ 
\end{tabular}
\end{center}
\end{ruledtabular}
\end{table*}

\begin{figure}[htbp]
\begin{center}
\includegraphics[scale=.8]{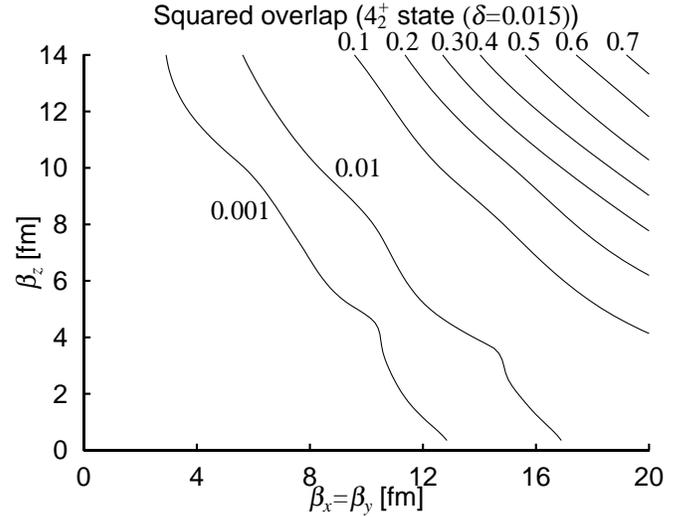}
\caption{Contour map of the squared overlap, $| \langle {\widehat \Phi}_{3\alpha}^{{\rm N},J=4}(\vc{\beta}) | \Psi _{\lambda=2}^{J=4}$($\delta$$=$$0.015) \rangle |^2$, in the two-parameter space, $\beta_x(=\beta_y)$ and $\beta_z$. Volkov No. 2 force is adopted.}\label{fig:42+}
\end{center}
\end{figure}

The fact that the r.m.s radius of the calculated $2_2^+$ state has the extraordinary large value of $5.40$ fm deserves special attention. It means that this state has about ten times the volume of the $^{12}$C ground state. One may, therefore, consider it as an ``alpha halo state''. As we mentioned in subsection \ref{hill}, we will show in our next paper that our $2_2^+$ wave function is very similar to that of Ref. \cite{kamimura} by Kamimura.

 Finally we study the $4_2^+$ state obtained in the previous subsection. We estimate the $4_2^+$ wave function by using the energy eigen functions on the $4_2^+$ trajectory shown in FIG. \ref{fig:4(c)}, as was done for the $2_2^+$ state. FIG. \ref{fig:42+} shows the squared overlap, $| \langle {\widehat \Phi}_{3\alpha}^{{\rm N},J=4}(\vc{\beta}) | \Psi _{\lambda=2}^{J=4}(\delta=0.015) \rangle |^2$. In spite of the fact that FIG. \ref{fig:4(c)} shows that the energy eigen function, $\Psi _{\lambda=2}^{J=4}(\delta=0.015)$ is sufficiently far from the nearest crossing point with the continuum state, the calculated squared overlap never has any non-negligible amplitudes in the inner interaction region. The root mean square radius of this state is 19.9 fm, and this state can hardly be regarded as a physical resonance state. If we make the coupling constant $\delta$ smaller than this value of $0.015$, going towards $\delta=0$, the wave function on this $4_2^+$ trajectory will have a larger radius than this unphysical value of $19.9$ fm. That is why we hesitate to suppose that the obtained energy position and width of the $4_2^+$ state reasonably correspond to a physical situation, although the obtained energy position for the effective nucleon force, Volkov No. 1, is comparable with the one by Uegaki {\it et al.}.

\subsection{The $0_3^+$ and $0_4^+$ states}\label{subsec:0304+}

\begin{figure}[htbp]
\begin{center}
\includegraphics[scale=0.7]{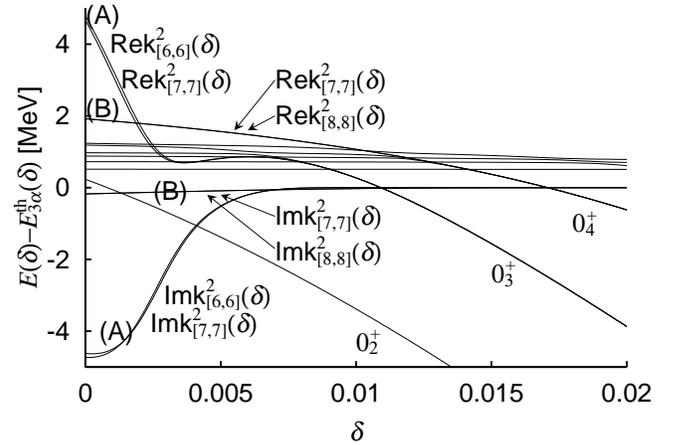}
\caption{Analytically continued complex energy functions for the $0_3^+$ and $0_4^+$ states are drawn by superposing them to FIG. \ref{fig:4(a)}. (A) is the analytically continued complex energy function for the $0_3^+$ state, and (B) for the $0_4^+$ state. Note that $k^2_{[6,6]}$ and $k^2_{[7,7]}$ for the $0_3^+$ state and  $k^2_{[6,6]}$ and $k^2_{[7,7]}$ for the $0_4^+$ state are converged, so that the two trajectories in each case of the $0_3^+$ and $0_4^+$ states are hardly distinguishable in this figure. Volkov No. 2 force is adopted.}\label{fig:5(a)}

\end{center}
\end{figure}

As we see in FIGs. \ref{fig:4(a)}-\ref{fig:4(c)} shown in the previous section, \ref{subsec:resaccc}, it turns out that there exist several resonance states going under the threshold as $\delta$ increases. For the $0^+$ and $2^+$ states we see three resonance states and for the $4^+$ state we see two, up to $\delta=0.02$. For the $0^+$ state, the $0_3^+$ and $0_4^+$ trajectories seem to cross mutually in the region with $\delta$ smaller than $0.01$. In FIG. \ref{fig:5(a)} we show the complex energy functions analytically continued for both $0_3^+$ and $0_4^+$ states in the case of Volkov No. 2 force. Throughout this subsection we will refer to analytically continued energy functions of the $0_3^+$ and $0_4^+$ states as (A) and (B), respectively. 

We see that the trajectories of the real parts of (A) and (B) cross each other around $\delta=0.002$ before reaching $\delta=0$. And we also see that the trajectory of the real part of (A) is not straight but quite distorted in the region, $0.003 \leq \delta \leq 0.006 $. This unexpected situation of the crossing of two lines might cause the distortion for the trajectory of the real part of (A). For (A), ${\rm Re}(k^2_{[7,7]}(\delta=0))=4.8$ MeV and $\Gamma=-2{\rm Im}(k^2_{[7,7]}(\delta=0))=9.2$ MeV, and for (B), ${\rm Re}(k^2_{[8,8]}(\delta=0))=1.9$ MeV and $\Gamma=-2{\rm Im}(k^2_{[8,8]}(\delta=0))=0.36$ MeV. This kind of situation including the non-straight shape of the real part of (A) and the crossing of the real parts of (A) and (B) similarly occurs in the case of the Volkov force No. 1. In this case, the complex energy function corresponding to (A) gives ${\rm Re}(k^2_{[9,9]}(\delta=0))=6.9$ MeV and $\Gamma=-2{\rm Im}(k^2_{[9,9]}(\delta=0))=6.2$ MeV. Both $k^2_{[7,7]}(\delta)$ and $k^2_{[8,8]}(\delta)$ give almost the same trajectory as $k^2_{[9,9]}(\delta)$ in a sufficiently wide energy region. And the complex energy function corresponding to (B) gives ${\rm Re}(k^2_{[8,8]}(\delta=0))=1.8$ MeV and $\Gamma=-2{\rm Im}(k^2_{[8,8]}(\delta=0))=0.4$ MeV. Again both $k^2_{[6,6]}(\delta)$ and $k^2_{[7,7]}(\delta)$ give almost the same trajectory as $k^2_{[8,8]}(\delta)$ in a sufficiently wide energy region. In order to determine the coefficients of all these Pad\'e approximants, we adopted the energy region from several tens of keV to around 3 MeV below the $3\alpha$ threshold. Since the observed binding energy measured from the $3\alpha$ threshold and total decay width of the $0_3^+$ state are 3.0 MeV and 3.0$\pm$0.7 MeV, respectively, neither (A) nor (B) reasonably explains the experimental data.

We wonder if the ACCC method where the analytic continuation is made using the Pad\'e approximant works well in the case where this kind of crossing occurs. Also the distortion of the trajectory of the real part of (A) makes us feel doubtful whether (A) describes the resonance reasonably well.

In section \ref{intro}, we referred to Refs. \cite{en'yo} and \cite{neff} in which the density distribution of the intrinsic state of the $0_3^+$ state is reported to have linear-chain-like configuration of $3\alpha$'s. If this result is reasonable, our present approach in which strongly prolate deformations with $\beta_x$$=$$\beta_y$$<$0.5 fm and $\beta_z$$>$3.0 fm are not included in the GCM calculation may not be suited for the description of the $0_3^+$ state. However it is not clear at all whether this possible inappropriateness of our approach is related or not to the crossing problem of the ACCC trajectories or the distorted shape of the ACCC trajectory of energy functions. Further studies are necessary to clear up the situation.

\section{Discussion}\label{sec:disc}

\subsection{Angular momentum component in the intrinsic wave function}\label{subsec:disc1}

\begin{figure}
\begin{center}
\subfigure[]{
\label{fig:amp0}
\includegraphics[scale=.85]{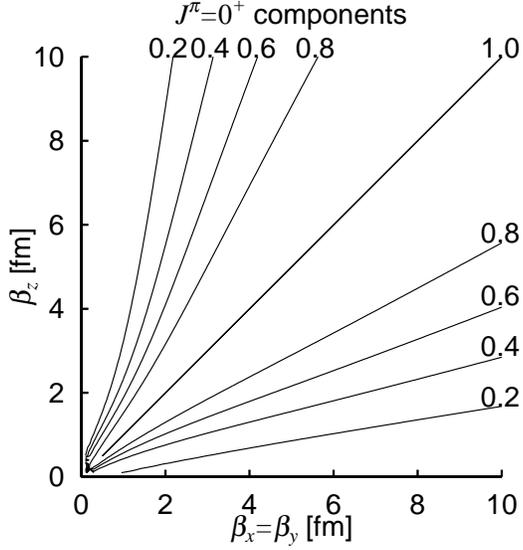}}
\subfigure[]{
\label{fig:amp2}
\includegraphics[scale=.85]{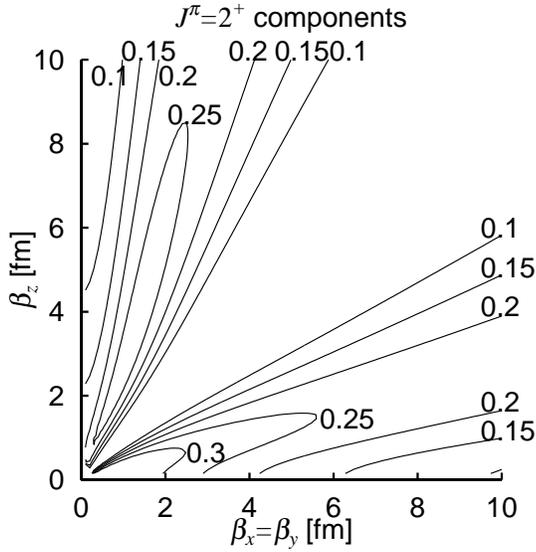}}
\caption{(a): Contour map of the squared amplitude, $|\langle \widehat{\Phi}^{{\rm N}, J=0}_{3\alpha}(\vc{\beta}) | \widehat{\Phi}^{\rm N}_{3\alpha}(\vc{\beta})\rangle |^2$ in the two-parameter space, $\beta_x(=\beta_y)$ and $\beta_z$. (b): Contour map of the squared amplitude, $|\langle \widehat{\Phi}^{{\rm N}, J=2}_{3\alpha}(\vc{\beta}) | \widehat{\Phi}^{\rm N}_{3\alpha}(\vc{\beta})\rangle |^2$ in the two-parameter space, $\beta_x(=\beta_y)$ and $\beta_z$. Harmonic oscillator size parameter $b$$=$$1.35$ fm is used in both figures.}
\end{center}
\end{figure}

Almost all of the studies of this paper are based on the wave function with good angular momentum quantum number, Eq. (\ref{eq:11}), which is projected out of the deformed intrinsic wave function, Eq. (\ref{eq:10}). The deformed wave function, $\widehat{\Phi}_{3\alpha}(\vc{\beta})$, is characterized as forming a gas-like $3\alpha$-cluster structure. However, in order to be able to impose the same character as the intrinsic wave function on the angular-momentum-projected wave function, it is necessary that the projected wave function is contained in the intrinsic wave function with a non-negligible amplitude. We define $\widehat{\Phi}^{\rm N}_{3\alpha}(\vc{\beta})$ as a normalized wave function of Eq. (\ref{eq:10}). For $J^\pi =0^+$ and $2^+$, we calculated the magnitude of the squared overlap amplitude, $|\langle \widehat{\Phi}^{{\rm N}, J}_{3\alpha}(\vc{\beta}) | \widehat{\Phi}^{\rm N}_{3\alpha}(\vc{\beta})\rangle |^2$, where $\widehat{\Phi}^{{\rm N}J}_{3\alpha}(\vc{\beta})$ is given in Eq. (\ref{eq:25}). The contour maps are shown in the FIGs. \ref{fig:amp0} and \ref{fig:amp2} for $J^\pi=0^+$ and $2^+$, respectively. The harmonic oscillator size parameter, $b=1.35$ fm is adopted. Even for the $2^+$ state, a non-negligible amplitude can be seen in both prolate and oblate deformed regions.

\subsection{Correlation between the $0^+_2$ and $0^+_1$ states}\label{subsec:disc2}

As already mentioned, the fact that the $0_2^+$ state has a dramatically different structure from the shell-model-like structure of the $0_1^+$ state was revealed by the works made about a quarter century ago \cite{uegaki, kamimura, hori}. There, the cluster structure of the $0_2^+$ state was characterized by the feature that the wave function has a large amplitude in the outer region, while the amplitude in the inner region is suppressed. The formation of the clustering structure of the $0_2^+$ state was attributed to the orthogonality to the $0_1^+$ wave function that has a compact structure, therefore, a large amplitude in the inner region. If this interpretation of the appearance of the cluster structure of the $0_2^+$ state is correct, the structure of the $0_2^+$ state would not be affected by the detailed structure of the $0_1^+$ state if only the $0_1^+$ state has a compact structure. 

\begin{figure}
\begin{center}
\includegraphics[scale=.85]{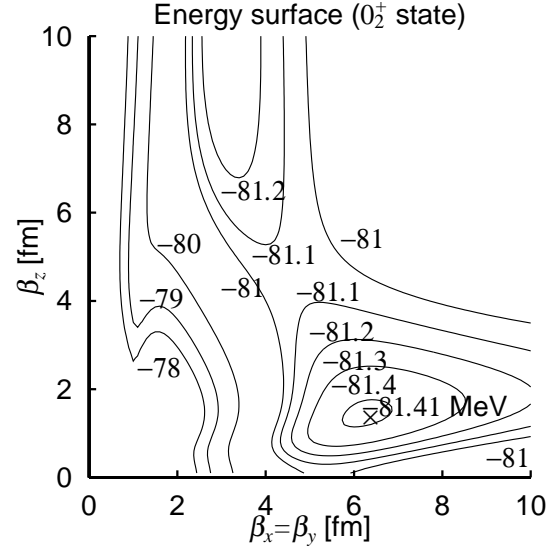}
\end{center}
\caption{Contour map of the energy surface corresponding to the $0^+$ 
state orthogonalized to the state at $\vc{\beta_0}^\prime$$=$($\beta_x$$=$$\beta_y$$=$$1.0$ fm, $\beta_z$$=$$2.0$ fm). The adopted effective nucleon force is Volkov No. 2. Numbers attached to the contour lines are binding energy values given in unit of MeV.}
\label{fig:independence1}
\end{figure}

In this subsection, we show quantitatively that the detailed structure of the $0_1^+$ state does not affect the structure of the $0_2^+$ state that is orthogonal to the $0_1^+$ state. In FIG. 1 of Ref. \cite{cbec}, we have shown the contour map of the energy surface of the $0^+$ state in the two-parameter space, $\beta_x(=\beta_y)$ and $\beta_z$, obtained by adopting the Volkov No. 2 force, which showed the minimum energy state at $\beta_x(=\beta_y)=1.5$ fm, $\beta_z=1.5$ fm, i.e. $\vc{\beta_0} \equiv (1.5\ {\rm fm},\ 1.5\ {\rm fm})$. The obtained minimum energy was $-87.68$ MeV, which was considered to correspond to the approximate binding energy of the $0_1^+$ state. In the reference paper, we further introduced the projection operator, $P^{J=0}_\bot (\vc{\beta_0})$ which is shown in Eq. (\ref{eq:261}) of this paper, and then looked for the minimum energy state for the $0^+$ state in the orthogonalized space, $P^{J=0}_\bot (\vc{\beta_0})\widehat{\Phi}_{3\alpha}^{J=0}(\vc{\beta})$. The minimum energy state was obtained at $\beta_x(=\beta_y)=5.7$ fm and $\beta_z=1.3$ fm, and the minimum energy was $-81.55$ MeV, which was considered to correspond to the approximate binding energy of the $0_2^+$ state. 

 In order to study the sensitivity of the $0_2^+$ state on the $0_1^+$ structure quantitatively, we introduce $\vc{\beta_0}^\prime \equiv (1.0\ {\rm fm},\ 2.0\ {\rm fm})$ and calculate the energy surface using the orthogonalized wave function $P_\bot^{J=0} (\vc{\beta_0}^\prime)\widehat{\Phi}_{3\alpha}^{J=0}(\vc{\beta})$. The binding energy of the wave function, $\widehat{\Phi}_{3\alpha}^{J=0}(\vc{\beta_0}^\prime)$ is calculated as $-85.96$ MeV, using the Volkov force, No. 2, which is higher by $1.72$ MeV than the value corresponding to $\widehat{\Phi}_{3\alpha}^{J=0}(\vc{\beta_0})$.

FIG. \ref{fig:independence1} is the contour map of the energy surface corresponding to the $0^+$ state, in the orthogonalized space, $P^{J=0}_\bot(\vc{\beta_0}^\prime)\widehat{\Phi}^{J=0}_{3\alpha}(\vc{\beta})$. We can see that there appears the minimum energy point at $\beta_x$$=$$\beta_y$$=$$6.2$ fm, $\beta_z$$=$$1.4$ fm with the minimum energy $-81.41$ MeV and the qualitative feature is quite similar to FIG. 2 of Ref. \cite{cbec}. In spite of the fact that the binding energy given by $\widehat{\Phi}_{3\alpha}^{J=0}(\vc{\beta_0}^\prime)$ is higher than the value given by $\widehat{\Phi}_{3\alpha}^{J=0}(\vc{\beta_0})$ by $1.72$ MeV, the minimum energy and the minimum position in FIG. \ref{fig:independence1}, are almost unchanged compared with those calculated using $P^{J=0}_\bot (\vc{\beta_0})\widehat{\Phi}_{3\alpha}^{J=0}(\vc{\beta})$. This result is consistent with the argument qualitatively made by the previous works referred to above, and further assures that the $0_2^+$ state could be separated from the $0_1^+$ state and treated independently under the condition that the state cannot become compact like the $0_1^+$ state.
\\

\subsection{Dominance of spherical component in the condensate wave function}\label{sph}

 If our GCM wave function of the $0^+_2$ state, $\Psi^{J=0}_{\lambda=2}$ is extended to a deformed type of the $\alpha$ condensate, one may ask the question how important this effect is with respect to the spherical case ? 

The discussion of the \ref{subsec:analysis} subsection teaches us that the deformed condensate wave function, after angular momentum projection, necessarily contains a large amount of the spherical condensate component. Actually we showed that after the projection onto the $J^\pi=0^+$ state, both prolate and oblate condensate wave functions take the form of Eq. (\ref{eq:39}) for the leading term which is nothing but the spherical condensate wave function.

In order to check the above argument quantitatively, we investigated the magnitude of the spherical condensate component which incorporates $\Psi^{J=0}_{\lambda=2}$. Let ${\cal V}_{\rm sph}$ and ${\cal W}_{\rm def}$ be the functional spaces spanned by spherical and by deformed condensate wave functions, respectively; ${\cal V}_{\rm sph} \subseteq {\cal W}_{\rm def}$, and $^\bot{\cal V}_{\rm sph}$ be the orthogonal functional space to ${\cal V}_{\rm sph}$. The projection operator $P_{\rm sph}$ onto ${\cal V}_{\rm sph}$ is written as $P_{\rm sph} = \sum_n |\Psi^n_{\rm sph} \rangle \langle \Psi^n_{\rm sph} |$, where $\Psi^n_{\rm sph}$ are the orthonormal basis functions of the space ${\cal V}_{\rm sph}$. The $\Psi^n_{\rm sph}$ are constructed as follows,
\begin{eqnarray*}
 && \sum_{\beta_x} \langle {\widehat \Phi}_{3\alpha}^{J=0}(\beta_x^\prime=\beta_y^\prime=
 \beta_z^\prime) | {\widehat \Phi}_{3\alpha}^{J=0}(\beta_x= \beta_y= \beta_z)
 \rangle g^n(\beta_x) \\ 
 && \hspace{6cm}= \mu_n g^n(\beta_x^\prime), \\ 
 && \sum_{\beta_x} g^m(\beta_x) g^n(\beta_x)=\delta_{mn}, \\
 && \Psi^n_{\rm sph} =\frac{1}{\sqrt{\mu_n}} \sum_{\beta_x} g^n(\beta_x) 
 {\widehat \Phi}_{3\alpha}^{J=0}(\beta_x= \beta_y= \beta_z).
\end{eqnarray*}
$\Psi^{J=0}_{\lambda=2}$ can be expanded by a linear combination of $\Psi^{n=1}_{\rm sph},\Psi^{n=2}_{\rm sph} \cdots \in {\cal V}_{\rm sph}$ and a vector $\chi \in \ ^\bot{\cal V}_{\rm sph} \cap {\cal W}_{\rm def}$ as
$$
\Psi^{{\rm N},J=0}_{\lambda=2} = c_1 \Psi^{n=1}_{\rm sph} + c_2 \Psi^{n=2}_{\rm sph}+ \cdots + \chi,
$$
where the relation, $P_{\rm sph} \chi =0$ is satisfied. 
The spherical condensate component can be calculated as $\langle \Psi^{{\rm N},J=0}_{\lambda=2}|P_{\rm sph}|\Psi^{{\rm N},J=0}_{\lambda=2}\rangle$. When the number of the adopted components $\Psi^n_{\rm sph}$ in $P_{\rm sph}$ is 16, where $\beta_x=\beta_y=\beta_z$ varies from $0.5$ fm to $15.5$ fm by 1 fm steps, the obtained values of the spherical condensate component are 92.1 \% and 90.6 \% for the Volkov force, No. 1 and No. 2, respectively. Even when we decreased the number of the adopted components to 13 from 16, i.e. $\beta_x=\beta_y=\beta_z=0.5$ fm to 12.5 fm by 1 fm steps, there appeared no difference of the obtained values at least to six significant figures for both forces. From this result it follows that our obtained values have already converged. The large magnitudes of these values imply that $\Psi^{J=0}_{\lambda=2}$ is composed of the spherical condensate component by more than 90 \%. At the same time we have to note that some amount (less than 10 \%) of the deformed component orthogonal to the spherical one is necessary in order to have quantitatively good reproduction of the observed results.

\subsection{Vibrational structure of the angular momentum projected wave function of the deformed condensate}

The expressions of the leading terms of the angular momentum projected wave functions of the deformed condensate, Eq. (\ref{eq:39}) for the $0^+$ state and Eq. (\ref{eq:40}) for the $2^+$ state, can be rewritten as follows,
\begin{eqnarray}
{\cal A} \Big[ \exp \Big\{- \frac{2}{B^2} \sum_{i=1}^{n-1} \mu_i {\vc \xi}_i^2 \Big\} \phi^n(\alpha) \Big] && \nonumber \\
&& \hspace{-5cm} = {\cal A} \Big[ \exp \Big\{- \frac{2}{B^2} \sum_{i=1}^{n} \Bigl( \vc{X}_i - \vc{X}_G \Bigr)^2 \Big\} \phi^n(\alpha) \Big] ,\label{eq:f1} \\
&& \hspace{-5.6cm}\frac{1}{B^2} \equiv \frac{2}{3B_x^2}+\frac{1}{3B_z^2},\nonumber
\end{eqnarray}
for the $0^+$ state, and
\begin{eqnarray}
&&{\cal A} \Big[ \exp \Big\{- \frac{2}{B^2} \sum_{i=1}^{n-1} \mu_i {\vc \xi}_i^2 \Big\} \sum_{i=1}^{n-1} \mu_i {\vc \xi}_i^2 Y_{20}(\vc{\widehat{\xi}}_i) \phi^n(\alpha) \Big] \nonumber \\
&& = {\cal A} \Big[ \exp \Big\{- \frac{2}{B^2} \sum_{i=1}^{n} \Bigl( \vc{X}_i - \vc{X}_G \Bigr)^2 \Big\}\nonumber \\ 
&& \hspace{1cm}\times \sum_{i=1}^n (\vc{X}_i-\vc{X}_G)^2 Y_{20}(\vc{\widehat{X_i-X_G}}_i) \phi^n(\alpha) \Big]\nonumber \\
&& = \sum_{j=1}^n {\cal A} \Big[ (\vc{X}_j-\vc{X}_G)^2 Y_{20}(\vc{\widehat{X_i-X_G}}_i)\nonumber \\
&& \hspace{1cm}\times \prod_{i=1}^n \exp \Big\{ -\frac{2}{B^2}(\vc{X}_i-\vc{X}_G)^2 \Big\}\phi(\alpha_i) \Big], \label{eq:f2}
\end{eqnarray}
for the $2^+$ state. The functional form of Eq. (\ref{eq:f1}) means, as was mentioned several times, that all $n$ $\alpha$-particles occupy an identical $S$-orbit, $\exp \{-(2/B^2)(\vc{X}-\vc{X}_G)^2\} $. On the other hand, the functional form of Eq. (\ref{eq:f2}) means that one $\alpha$-particle occupies a $D$-orbit, $(\vc{X}-\vc{X}_G)^2 Y_{20}(\vc{X}-\vc{X}_G)\exp \{-(2/B^2)(\vc{X}-\vc{X}_G)^2\} $ while the other $(n-1)$ $\alpha$-particles occupy an identical $S$-orbit, $\exp \{-(2/B^2)(\vc{X}-\vc{X}_G)^2\}$. Therefore, our $2^+$ wave function constructed by the angular momentum projection of the deformed condensate wave function has the structure that one $\alpha$-particle is excited from the $S$-orbit into a $D$-orbit while the other $(n-1)$ $\alpha$-particles remain in the identical $S$-orbit which is occupied by all $n$ $\alpha$-particles in the case of the $0^+$ state.

 We can yet give another explanation for the $2^+$ wave function and this explanation can be applied to any wave function with spin obtained by the angular momentum projection of the deformed condensate wave function. We rewrite Eq. (\ref{eq:38}) as follows,
\begin{eqnarray}
&&\exp \Big\{ -2\sum_{i=1}^{n-1}\mu_i \Big( \frac{\xi_{ix}^2+\xi_{iy}^2}{B_x^2}+\frac{\xi_{iz}^2}{B_z^2}\Big)\Big\} \nonumber \\
&&\hspace{0.5cm}=\exp\Big\{ -\frac{2}{B^2}\sum_{i=1}^n(\vc{X}_i-\vc{X}_G)^2 \Big\}\nonumber \\
&& \hspace{1.cm} \times \Big\{ 1+ \sigma \widehat{D} + \frac{\sigma^2}{2 !}\widehat{D}^2 +\frac{\sigma^3}{3 !}\widehat{D}^3 + \cdots \Big\}, \\
&& \widehat{D}=\sum_{i=1}^n (\vc{X}_i-\vc{X}_G)^2Y_{20}(\widehat{\vc{X}_i-\vc{X}_G}),\nonumber \\
&& \sigma=-\frac{2}{3}\sqrt{\frac{16\pi}{5}}\Big( \frac{1}{B_x^2}-\frac{1}{B_z^2} \Big).\nonumber
\end{eqnarray}
We can say that the leading term of the $2^+$ state is created from the $0^+$ state by one $D$-phonon-like excitation expressed by the operator $\widehat{D}$. Similarly, the leading term of the $4^+$ state is created by two $D$-phonon-like excitation expressed by the operator $(\widehat{D}^2)_{J=4}$, where $(\widehat{D}^2)_{J=4}$ stands for the operator obtained from $\widehat{D}^2$ by the angular momentum projection onto $J=4$. We can easily see that any spin $J$ wave function obtained from the deformed condensate wave function has the leading term which has the form of multiple $D$-phonon-like excitation from the $0^+$ state expressed by the operator $(\widehat{D}^m)_J$. 

 The above argument shows that all the angular momentum projected wave functions of the deformed condensate can be viewed to express the state of vibration-like excitation rather than the rotational excitation. However, the GCM calculation by superposing the spin-projected condensate wave function is not directly related to the diagonalization of the Hamiltonian whithin the multi-phonon-like subspace with definite $J$, because in the GCM calculation we superpose phonon-like states with different oscillator parameters $B$.

\section{Conclusion}\label{sec:conc}

We investigated the states with $J^\pi$$=$$0^+$, $2^+$ and $4^+$, of $^{12}$C with the excitation energy less than about $15$ MeV by using the $3\alpha$ condensate wave function with spatial deformation. In order to discuss the total alpha decay widths as well as the energy positions we applied the ACCC method to the wave function. 

We could successfully reproduce both the binding energy and total alpha decay width of the recently observed $2_2^+$ state. The observed values of the energy and width are 2.6 $\pm$ 0.3 MeV and 1.0 $\pm$ 0.3 MeV, respectively. The calculated binding energy and width are 2.1 MeV and 0.64 MeV, respectively, for the use of Volkov No. 1 force. For the use of Volkov No. 2 force, when the binding energy is adjusted to 2.0 MeV with the nuclear interaction being a little weakened, the width is calculated as 0.61 MeV. We found that in the case where the binding energy of the $2_2^+$ state is adjusted to 0.26 MeV which is the calculated binding energy of the $0_2^+$, the density distribution of the $2_2^+$ state is almost the same as that of the $0_2^+$ state for the use of Volkov No. 2 force. And then, the kinetic energy, nuclear interaction energy, and Coulomb interaction energy of the calculated $2_2^+$ state are found to be very similar to those of the $0_2^+$ state. 

We devised a new technique to extract the proper resonance component from the wave function with some mixture of continuum states which are obtained within the bound state approximation. This technique allows us to obtain pure resonance wave functions very easily within the framework of the bound state approximation. Detailed discussions about this new method will be given in our forthcoming paper. As a result of applying this method to our present problem, we found that the pure resonance $2_2^+$ wave function which gives almost the same resonance energy as the above mentioned ACCC result has a large overlap with the single condensate wave function of $3\alpha$ gas-like structure. The squared overlap value is more than 88 \% when we employ the Volkov No. 2 force.

These facts imply that the $2_2^+$ state has a similar structure to that of the $0_2^+$ state which has a gas-like structure composed of three alpha clusters and is of Bose-condensate character. The $2_2^+$ state is obtained by promoting just one alpha cluster out of the condensate of the $0_2^+$ state into a $D$-wave. The result of the present paper about the $2_2^+$ state of $^{12}$C suggests that for more heavier nuclei, for instance, $^{16}$O, there would exist non-zero spin excited states which are members of an alpha cluster condensate family. The $0_2^+$ state is a state of very dilute density which is about $1/3$ of the ground state density. The density of the $2_2^+$ state is pointed out to be even more dilute than that of the $0_2^+$ state. The $2_2^+$ state has an extraordinarily large r.m.s radius such that the volume of the $2_2^+$ state is a factor ten larger than the one of the ground state. We call such a wide $\alpha$-configuration an $\alpha$-halo state. These dilute densities or large radii of the $0_2^+$ and $2_2^+$ states are probably the reason why no-core shell model calculations of $^{12}$C completely fail to account for the $0_2^+$ and $2_2^+$ states in spite of their good reproduction of the ground state band \cite{nocore}. All these above features hint to the possibility of a description of these states in terms of ideal bosons as this has in deed been considered in \cite{yamada}.

We failed to give a clear interpretation of the $0_3^+$ state. As mentioned in the text, two recent works \cite{en'yo, neff} obtained a practically identical structure for the $0_3^+$ state consisting out of a $^8$Be-like entity with a third $\alpha$ attached to it by about $30^\circ$ off the symmetry axis of the first two $\alpha$'s. Though these calculations fail to reproduce the $0_3^+$ position by several MeV, such results are nevertheless often indicative of the geometrical structure of the state as this is also the case for the $0_2^+$ state in $^{12}$C obtained from an AMD aproach \cite{en'yo}. In these calculations, the $0_3^+$ state looks similar to a chain state but not quite. May be a very prolate deformed $\alpha$-condensate wave function could elucidate the physical interpretation of the state. It is our next task to investigate the $0_3^+$ state by incorporating strongly prolate condensed wave functions in the GCM calculation.

%with $\beta_x$$=$$\beta_y$$<$0.5 fm in the GCM calculation.

\begin{acknowledgements}
One of the authors (Y. F) highly appreciates many helpful comments and words of advice for the method of ACCC of Prof. S. Aoyama and Dr. T. Myo. He also would like to thank Ms. C. Kurokawa for fruitful discussions. Valuable comments of Prof. K. Kat${\rm \bar{o}}$ and Prof. Y. Fujiwara are also acknowledged. This work was partially performed in the Research Project for Study of Unstable Nuclei from Nuclear Cluster Aspects sponsored by Institute of Physical and Chemical Research (RIKEN), and is supported by the Grant-in-Aid for the 21st Century COE "Center for Diversity and Universality in Physics" from the Ministry of Education, Culture, Sports, Science and Technology (MEXT) of Japan.
\end{acknowledgements}

\appendix*
\section{Matrix elemelnts of density operator}

We explain how to calculate the matrix elements of the density operator defined in Eq. (\ref{eq:dsty}). Since the density operator maintains the rotational invariance, the expectation value by the normalized wave function with good angular momentum can be written as follows;
%\begin{widetext}
\begin{eqnarray}
&&\Big\langle {\widehat \Phi}_{n\alpha}^{{\rm N}J}(\vc{\beta}) \Big| {\widehat \rho}(a) \Big| {\widehat \Phi}_{n\alpha}^{{\rm N}J}(\vc{\beta^\prime}) \Big\rangle \nonumber \\
= && \frac{\Big\langle {\widehat \Phi}_{n\alpha}^J(\vc{\beta}) \Big| {\widehat \rho}(a) \Big| {\widehat \Phi}_{n\alpha}^J(\vc{\beta^\prime}) \Big\rangle}{\sqrt{\Big\langle {\widehat \Phi}_{n\alpha}^J(\vc{\beta}) \Big| {\widehat \Phi}_{n\alpha}^J(\vc{\beta}) \Big\rangle \Big\langle {\widehat \Phi}_{n\alpha}^J(\vc{\beta}^\prime) \Big| {\widehat \Phi}_{n\alpha}^J(\vc{\beta^\prime}) \Big\rangle }} \nonumber \\ 
 = && \frac{\Big\langle {\widehat \Phi}_{n\alpha}(\vc{\beta}) \Big| {\widehat \rho}(a) \Big| {\widehat \Phi}_{n\alpha}^J(\vc{\beta^\prime}) \Big\rangle}{\sqrt{\Big\langle {\widehat \Phi}_{n\alpha}(\vc{\beta}) \Big| {\widehat \Phi}_{n\alpha}^J(\vc{\beta}) \Big\rangle \Big\langle {\widehat \Phi}_{n\alpha}(\vc{\beta}^\prime) \Big| {\widehat \Phi}_{n\alpha}^J(\vc{\beta^\prime}) \Big\rangle}} . \nonumber \\
% &&= \int d \cos\theta d^J_{00}(\theta) \Big\langle {\widehat \Phi}_{n\alpha}(\vc{\beta}) \Big|{\widehat \rho}(a) {\widehat R}_y(\theta) \Big| {\widehat \Phi}_{n\alpha}(\vc{\beta^\prime}) \Big\rangle . \nonumber \\ 
 && \label{ap:1}
\end{eqnarray}
%\end{widetext}
In order to evaluate the above expression, we make use of a similar technique to the one used in Eq. (\ref{eq:12}). At the final step of Eq. (\ref{eq:12}), we change the numerator as follows, keeping the denominator unchanged:
\begin{eqnarray}
&& \int d\cos\theta d^J_{00}(\theta) \frac{1}{P_0(\theta)}\Big\langle \Phi_{n\alpha}(\vc{\beta}) \Big|{\widehat \rho}(a){\widehat R}_y(\theta) \Big| \Phi_{n\alpha}(\vc{\beta^\prime}) \Big\rangle  \nonumber \\
&& = \int d\cos\theta d^J_{00}(\theta) \frac{\Big\langle \Phi_{n\alpha}(\vc{\beta}) \Big|\delta(\vc{X}_G) {\widehat \rho}(a) {\widehat R}_y(\theta) \Big| \Phi_{n\alpha}(\vc{\beta^\prime})\Big\rangle}{\Big\langle \Phi^{\vc{X}_G}_{n\alpha}(\vc{\beta}) \Big|\delta(\vc{X}_G)  {\widehat R}_y(\theta)\Big|\Phi^{\vc{X}_G}_{n\alpha}(\vc{\beta}^\prime) \Big\rangle} \nonumber \\
&& = \int d\cos\theta d^J_{00}(\theta) \Big\langle \Phi_{n\alpha}(\vc{\beta}) \Big|\delta(\vc{X}_G) {\widehat \rho}(a) {\widehat R}_y(\theta) \Big| \Phi_{n\alpha}(\vc{\beta^\prime})\Big\rangle .\nonumber \\
&& \label{ap:2}
\end{eqnarray}
Here, we made use of the relation of Eq. (\ref{eq:10a})
\begin{equation}
\Phi_{n\alpha}(\vc{\beta})=Q\ \Phi^{\vc{X}_G}_{n\alpha}(\vc{\beta})\ {\widehat \Phi}_{n\alpha}(\vc{\beta}),
\end{equation}
where $Q$ is a constant, $\Phi^{\vc{X}_G}_{n\alpha}(\vc{\beta})$ is defined as 
\begin{equation}
\Phi^{\vc{X}_G}_{n\alpha}(\vc{\beta})
\equiv \exp \Bigl( - \sum_{k=x,y,z} 
 \frac{2n}{B_k^2} X_{Gk}^2 \Bigr), \label{ap:3}
\end{equation}
and use is made of the trivial relation,
\begin{equation}
\Big\langle \Phi^{\vc{X}_G}_{n\alpha}(\vc{\beta}) \Big|\delta(\vc{X}_G)  {\widehat R}_y(\theta)\Big|\Phi^{\vc{X}_G}_{n\alpha}(\vc{\beta}^\prime) \Big\rangle =1.  \label{ap:4}
\end{equation}
Next, by using the following relation, 
\begin{equation}
\int d^2\vc{\widehat a} \sum_{lm} Y^\ast _{lm}(\widehat{\vc{r}_i-\vc{X}_G})Y_{lm}(\vc{\widehat a})=1, \label{ap:5}
\end{equation}
the density operator is represented below as
\begin{eqnarray}
{\widehat \rho}(a)&=&\frac{1}{A} \sum_{i=1}^A \delta (|\vc{r}_i-\vc{X}_G|-a)\nonumber \\ 
&=&\frac{a^2}{A} \sum_{i=1}^A \int d^2 \vc{\widehat a}\ \delta(\vc{r}_i-\vc{X}_G-\vc{a}). \label{ap:6}
\end{eqnarray}
Here $\widehat{\vc{a}}$ and $\widehat{\vc{r}_i-\vc{X}_G}$ are polar angles of $\vc{a}$ and $\vc{r}_i-\vc{X}_G$, respectively. Substituting Eq. (\ref{ap:6}) into the final expression of Eq. (\ref{ap:2}), we only have to evaluate the following expression, while the integral over the variables, $\cos \theta$ in Eq. (\ref{ap:2}) and $\widehat{\vc{a}}$ in Eq. (\ref{ap:6}), is numerically performed afterwards.
\begin{eqnarray}
&&\Big\langle \Phi_{n\alpha}(\vc{\beta}) \Big|\ \delta(\vc{X}_G)\sum_{i=1}^A \delta(\vc{r}_i-\vc{X}_G-\vc{a}){\widehat R}_y(\theta)
 \Big| \Phi_{n\alpha}(\vc{\beta^\prime}) \Big\rangle \nonumber \\
&&\!\!\!\!= \Big\langle \Phi_{n\alpha}(\vc{\beta}) \Big|\ \delta(\vc{X}_G)\sum_{i=1}^A \delta(\vc{r}_i-\vc{a}){\widehat R}_y(\theta)
 \Big| \Phi_{n\alpha}(\vc{\beta^\prime}) \Big\rangle \nonumber \\
&&\!\!\!\!= \!\!\int\!\! \frac{d^3 k}{(2\pi)^3} \Big\langle \Phi_{n\alpha}(\vc{\beta}) \Big|\ e^{i\vc{k}\cdot\vc{X}_G}\sum_{i=1}^A \delta(\vc{r}_i-\vc{a}) {\widehat R}_y(\theta) \Big| \Phi_{n\alpha}(\vc{\beta^\prime}) \Big\rangle . \nonumber \\
&& \label{ap:7}
\end{eqnarray}
Considering the fact that the wave function, $\Phi_{n\alpha}(\vc{\beta})$ is represented by using Brink's wave function, as 

\begin{eqnarray}
&&{\widehat R}_y(\theta)\Big|\Phi_{n\alpha}(\vc{\beta})\Big\rangle = \int d^3 R_1\ \cdots d^3 R_n \times \nonumber \\ 
&& \exp \Bigl( -\sum_{i=1}^n \sum_{k=x,y,z} \frac{R_{ik}^2}{\beta_k^2} \Bigr) \Big|\Phi^{\rm B}({\widehat R}^{-1}_y(\theta){\vc R}_1, \cdots,  R^{-1}_y(\theta){\vc R}_n)\Big\rangle \nonumber \\ 
 && = \int d^3 R_1\ \cdots d^3 R_n \times \nonumber \\ 
&& \exp \Big\{ -\sum_{i=1}^n \sum_{k=x,y,z} \frac{({\widehat R}_y(\theta){\vc R}_i)_k^2} {\beta_k^2} \Big\} \Big|\Phi^{\rm B}({\vc R}_1, \cdots, {\vc R}_n)\Big\rangle, \nonumber \\
&& \label{ap:8}
\end{eqnarray}
we should be able to evaluate the following quantity,
\begin{eqnarray}
&& \int \frac{d^3 k}{(2\pi)^3} \Big\langle \Phi^{\rm B}(\vc{R}_1, \cdots, \vc{R}_n) \Big|\ e^{i\vc{k}\cdot\vc{X}_G} \times \nonumber \\ 
&&\hspace{3cm} \sum_{i=1}^A \delta(\vc{r}_i-\vc{a}) \Big| \Phi^{\rm B}(\vc{R}^\prime_1, \cdots, \vc{R}^\prime_n) \Big\rangle \nonumber \\
&&= \int \frac{d^3 k}{(2\pi)^3} \Big\langle \Phi^{\rm B}(\vc{R}_1, \cdots, \vc{R}_n) \Big|\prod_{j=1}^A\ \exp(\frac{i}{A}\vc{k}\cdot \vc{r}_j) \times \nonumber \\ 
&&\hspace{3cm} \sum_{i=1}^A \delta(\vc{r}_i-\vc{a}) \Big| \Phi^{\rm B}(\vc{R}^\prime_1, \cdots, \vc{R}^\prime_n) \Big\rangle . \nonumber \\
&& \label{ap:9}
\end{eqnarray}
All of the following procedures are well known and can be straightforwardly performed, once a single particle wave function, $\varphi_i(\vc{r})$, and a modified single particle wave function, $\varphi_i^\prime(\vc{r})$, are introduced as follows,
\begin{eqnarray}
\varphi_i(\vc{r})       &=& (\pi b^2)^{-3/4}\exp\Big(-\frac{1}{2b^2}(\vc{r}-\vc{R}_i)^2 \Big), \nonumber \\
\varphi_i^\prime(\vc{r})&=& (\pi b^2)^{-3/4}\exp\Big(-\frac{1}{2b^2}(\vc{r}-\vc{R}^\prime_i)^2 + \frac{i}{A}\vc{k}\cdot \vc{r} \Big), \nonumber \\
&& \hspace{3cm} (i=1,\cdots, n) .\nonumber \\
&& \label{ap:10}
\end{eqnarray}
The overlap kernel between $\varphi_i(\vc{r})$ and $\varphi^\prime_j(\vc{r})$ is calculated as 
\begin{eqnarray}
\tilde{D}_{ij} & \equiv & \langle \varphi_i | \varphi^\prime_j \rangle \nonumber \\
&=& \exp\Big(-\frac{b^2\vc{k}^2}{4A^2} + \frac{i}{2A}\vc{k}\cdot(\vc{R}_i+\vc{R}^\prime_j) \Big) D_{ij} , \nonumber \\
&& \label{ap:11}
\end{eqnarray}
where we use a shorthand notation, $D_{ij}=\langle \varphi_i | \varphi_j \rangle$. $\tilde{D}_{ij}$ and $D_{ij}$ are defined as $n\times n$ matrices, and the inverse matrix of $\tilde{D}_{ij}$ is also calculated easily,
\begin{eqnarray}
(\tilde{D}^{-1})_{ij} & \equiv & \langle \varphi_i | \varphi^\prime_j \rangle ^{-1} \nonumber \\
&=& \exp\Big(\frac{b^2\vc{k}^2}{4A^2} - \frac{i}{2A}\vc{k}\cdot(\vc{R}^\prime_i+\vc{R}_j) \Big) (D^{-1})_{ij} . \nonumber \\
&& \label{ap:12}
\end{eqnarray}
Now we will see that using Eqs. (\ref{ap:11}), (\ref{ap:12}), Eq. (\ref{ap:9}) is represented as follows:
\begin{eqnarray}
&&\int \frac{d^3 k}{(2\pi)^3} \Big\langle \Phi^{\rm B}(\vc{R}_1, \cdots, \vc{R}_n) \Big|\prod_{j=1}^A\ \exp(\frac{i}{A}\vc{k}\cdot \vc{r}_j) \times \nonumber \\ 
&&\hspace{3cm} \sum_{i=1}^A \delta(\vc{r}_i-\vc{a}) \Big| \Phi^{\rm B}(\vc{R}^\prime_1, \cdots, \vc{R}^\prime_n) \Big\rangle \nonumber \\
&&=\int \frac{d^3 k}{(2\pi)^3} 4|\tilde{D}|^4\sum_{i,j}^n \langle \varphi_i | \delta(\vc{r}-\vc{a}) | \varphi_j^\prime \rangle (\tilde{D}^{-1})_{ji} . \nonumber \\
&& \label{ap:13}
\end{eqnarray}
After substituting the explicit formulae of $\tilde{D}_{ij}$ and $(\tilde{D}^{-1})_{ji}$ into the above expression and performing the Gauss integral over the variable, $\vc{k}$, analytically, we finally arrive at the following formula:
\begin{eqnarray}
&&\int \frac{d^3 k}{(2\pi)^3} 4|\tilde{D}|^4\sum_{i,j}^n \langle \varphi_i | \delta(\vc{r}-\vc{a}) | \varphi_j^\prime \rangle (\tilde{D}^{-1})_{ji} \nonumber \\
&&=\frac{4}{(2\pi)^3}\Big\{\frac{4A^2}{b^4(A-1)}\Big\}^{3/2} |D|^4\sum_{ij}^n D_{ij} (D^{-1})_{ji} \times \nonumber \\  
&& \exp\Big[ -\frac{1}{b^2}\Bigl(\vc{a}-\frac{\vc{R}_i+\vc{R}^\prime_j}{2}\Bigr)^2 \nonumber \\ 
&& -\frac{1}{b^2(A-1)}\Big\{ \vc{a}+2\sum_k^n(\vc{R}_k+\vc{R}^\prime_k) - \frac{\vc{R}_i+\vc{R}^\prime_j}{2} \Big\}^2  \Big].
\nonumber \\
&& \label{ap:14}
\end{eqnarray}
Note that it is not complicated to perform the Gaussian integral over the variables, $\vc{R}_1, \cdots, \vc{R}_n$. As mentioned above, the integral over the variables, $\vc{\widehat a}$ and $\cos \theta$, is performed numerically.

\bibliography{albec12ref.bib}
%\bibliographystyle{unsrt}

%@ARTICLE{feyn54,
%   author = "R. P. Feynman",
%   year = "1954",
%   journal = "Phys.\ Rev.",
%   volume = "94",
%   pages = "262"
%}

%@ARTICLE{epr,
%   author = "A. Einstein and B. Podolsky and N. Rosen",
%   journal = "Phys.\ Rev.",
%   volume = "47",
%   pages = "777"
%}

%@MISC{witten2001,
%   author = "Edward Witten",
%   eprint = "hep-th/0106109"
%}

\end{document}